\documentclass[reqno,11pt]{amsart}

\usepackage{multind}
\usepackage{graphicx}
\usepackage{amscd}
\usepackage{slashed}
\usepackage{amssymb}
\usepackage{esint}
\usepackage[mathscr]{eucal}
\numberwithin{equation}{section}
\allowdisplaybreaks[1]

\title[Perturbative Description of the Fermionic Projector]{Perturbative Description of the Fermionic Projector:
Normalization, Causality and Furry's Theorem}

\author[F.\ Finster]{Felix Finster}
\address{Fakult\"at f\"ur Mathematik \\ Universit\"at Regensburg \\ D-93040 Regensburg \\ Germany}
\email{finster@ur.de}

\author[J.\ Tolksdorf]{J\"urgen Tolksdorf \\ \\ January 2014}
\address{Max Planck Institute for Mathematics in the Sciences, Leipzig, Germany}
\email{Juergen.Tolksdorf@mis.mpg.de}
\thanks{The research leading to these results has received funding from the
European Research Council under the European Union's Seventh Framework
Programme (FP7/2007-2013) / ERC grant agreement n$^\circ$~267087.}

\newtheorem{Def}{Definition}[section]
\newtheorem{Thm}[Def]{Theorem}
\newtheorem{Prp}[Def]{Proposition}
\newtheorem{Lemma}[Def]{Lemma}

\newtheorem{Corollary}[Def]{Corollary}

\newcommand{\Thanks}{\vspace*{.5em} \noindent \thanks}
\newcommand{\beq}{\begin{equation}}
\newcommand{\eeq}{\end{equation}}
\newcommand{\Proof}{\begin{proof}}
\newcommand{\QED}{\end{proof} \noindent}

\newcommand{\la}{\langle}
\newcommand{\ra}{\rangle}
\newcommand{\bra}{\mathopen{<}}
\newcommand{\ket}{\mathclose{>}}

\newcommand{\R}{\mathbb{R}}
\newcommand{\1}{\mbox{\rm 1 \hspace{-1.05 em} 1}}
\newcommand{\Z}{\mathbb{Z}}

\newcommand{\Pdd}{\mbox{$\partial$ \hspace{-1.2 em} $/$}}

\renewcommand{\H}{\mathscr{H}}

\renewcommand{\O}{\mathscr{O}}

\newcommand{\B}{{\mathscr{B}}}
\renewcommand{\O}{{\mathscr{O}}}

\newcommand{\res}{\text{\rm{res}}}
\newcommand{\ret}{\wedge}
\newcommand{\sea}{\text{\rm{sea}}}

\DeclareMathOperator{\Tr}{Tr}

\DeclareMathOperator{\Pexp}{Pexp}
\DeclareMathOperator{\sdot}{\cdot}

\begin{document}
\maketitle

\begin{abstract}
The causal perturbation expansion of the fermionic projector is performed with a contour integral method.
Different normalization conditions
are analyzed. It is shown that the corresponding light-cone expansions are causal
in the sense that  they only involve bounded line integrals.
For the resulting loop diagrams we prove a generalized Furry theorem.
\end{abstract}

\tableofcontents


\section{Introduction}
The causal perturbation theory as developed in~\cite{sea, grotz} gives a perturbative description
of the Dirac sea in an external potential (see also~\cite[Chapter~2]{PFP}).
It is the starting point for a detailed analysis of the fermionic projector in position space~\cite{firstorder, light},
which forms the technical core of the fermionic projector approach to
quantum field theory (see~\cite{srev} and the references therein).
More recently, the reformulation in terms of causal fermion systems~\cite{rrev} and the
non-perturbative construction of the fermionic projector in~\cite{finite, infinite, hadamard} shed a new light on
how the fermionic projector should be normalized.
Moreover, the spectral methods used in the non-perturbative construction motivated
that the perturbation expansions should be described more efficiently with contour integrals.
Finally, the systematic treatment of perturbative quantum field theory in
the framework of the fermionic projector in~\cite{qft} showed that fermion loops are to be described in
a specific formalism involving integral kernels~$L_\ell$ to be formed of the contributions to the
perturbation expansion in an external potential. This raises the question which of these integral kernels
vanish in analogy to Furry's theorem in standard quantum field theory. The goal of the present paper is to
treat all these issues in a coherent and conceptually convincing way, also giving a systematic procedure for
all computations needed in future applications.

The paper is organized as follows. In Section~\ref{secnorm} we recall the definition of
the fermionic projector in the Minkowski vacuum and explain the different methods for
its normalization, referred to as the {\em{mass normalization}} and the {\em{spatial normalization}}.
In Section~\ref{seccontour} we perform the perturbation expansion with contour integral methods,
both with mass and spatial normalization.
In Section~\ref{secflow} the perturbation expansions are described by
the so-called unitary perturbation flow, which is particularly useful if particle and/or anti-particle
states are present.
In Section~\ref{secother} we analyze the retarded expansion and the expansion
with the Feynman propagator as alternative perturbation expansions.
Section~\ref{seclight} is devoted to the light-cone expansion of the resulting Feynman diagrams.
It is shown that the light-cone expansions of all diagrams is causal in the sense that it
only involves bounded line integrals.
In Section~\ref{secloop} we analyze the resulting loop diagrams and prove a generalized
Furry theorem which states that certain classes of loop diagrams vanish.
Finally, in Appendix~\ref{appA} we list the leading orders of the relevant perturbation expansions.

\section{The Normalization of the Vacuum Fermionic Projector} \label{secnorm}
We let~$(M, \la.,. \ra)$ be Minkowski space
(with the signature convention~$(+ \ \!\! - \ \!\! - \ \! - )$).
In the vacuum, a completely filled Dirac sea is described by the distribution
(for basics see~\cite[Chapter~2]{PFP} or~\cite{intro})
\beq \label{Fourier1}
P_m(k) = (\slashed{k}+m)\: \delta(k^2-m^2)\: \Theta(-k^0)
\eeq
(where~$\Theta$ denotes the Heaviside function).
Taking the Fourier transform, we obtain the distribution
\beq \label{Fourier2}
P_m(x,y) = \int \frac{d^4k}{(2 \pi)^4}\: P_m(k)\: e^{-ik(x-y)}\:,
\eeq
referred to as the {\em{kernel of the fermionic projector}} of the vacuum.
We also consider this distribution as the integral kernel of an operator acting on
wave functions in space-time, i.e.
\beq \label{Pmop}
(P_m \psi)(x) = \int_M P_m(x,y)\: \psi(y)\: d^4y \:.
\eeq
This operator is the so-called fermionic projector. A point causing major complications
is that for the space-time integral in~\eqref{Pmop} to exist, 
the wave function~$\psi$ must have suitable decay properties at infinity
(for example, \eqref{Pmop} is well-defined for test functions~$\psi \in C^\infty_0(M)$ as
the convolution by a distribution). In particular, the time integral in~\eqref{Pmop}
in general diverges if~$\psi$ is a physical wave function, being a solution of the Dirac equation.
A closely related problem is that one cannot multiply the operator~$P_m$ naively by itself
because the formal integral
\beq \label{PPint}
\int P_m(x,z)\, P_m(z,y) \,d^4y
\eeq
is ill-defined. This problem can be understood similarly in momentum space.
Namely, using that convolution in position space corresponds to multiplication in momentum space,
the integral in~\eqref{PPint} corresponds to the formal product
\beq \label{PmPm}
P_m(k)\, P_m(k) \:,
\eeq
which is again ill-defined because the square of the $\delta$-distribution in~\eqref{Fourier1}
makes no mathematical sense.
As we shall see, these obvious problems in the naive treatment of the fermionic projector are not
only a mathematical subtlety. On the contrary, the methods for overcoming these problems
will involve a careful analysis of the causal structure of the fermionic projector and of its
proper normalization.

\subsection{The Mass Normalization and the Spatial Normalization in the Vacuum}
The fermionic projector of the vacuum is normalized in two different ways.
First, considering the mass~$m$ as a variable parameter, one can
consider the product~\eqref{PmPm} with two different masses,
\begin{align*}
P_m(k) \: P_{m^\prime}(k) &= (\slashed{k} + m) \: \delta(k^2 - m^2) \:\Theta(-k^0)\;
   (\slashed{k} + m^\prime) \: \delta(k^2 - (m^\prime)^2) \:\Theta(-k^0) \\
&= \big( k^2 + (m+m^\prime)\, \slashed{k} + m m^\prime \big) \: \delta(m^2 -
   (m^\prime)^2) \; \delta(k^2 - m^2) \:\Theta(-k^0) \\
&= \big( k^2 + (m+m^\prime)\, \slashed{k} + m m^\prime \big) \: \frac{1}{2m} \:\delta(m -
   m^\prime) \; \delta(k^2 - m^2) \:\Theta(-k^0) \\
&= \delta(m-m^\prime) \: (\slashed{k} + m) \: \delta(k^2 - m^2) \:\Theta(-k^0)\: .
\end{align*}
We thus obtain the distributional identity
\beq \label{massnorm}
P_m \: P_{m^\prime} = \delta(m-m^\prime)\: P_m \:.
\eeq
This resembles idempotence, but it involves a $\delta$-distribution in the mass
parameter. We refer to~\eqref{massnorm} as the {\bf{mass normalization}}.

Alternatively, one can replace the space-time integral~\eqref{PPint} by
an integral only over space. This can be understood from the fact
that for a Dirac wave function~$\psi$, the quantity~$(\overline{\psi} \gamma^0 \psi)(t_0, \vec{x})$
has the interpretation as the probability density for the particle at time~$t_0$ to be at position~$\vec{x}$.
Integrating over space and polarizing, we obtain the scalar product
\beq \label{print}
(\psi | \phi)_{t_0} = \int_{\R^3} \overline{\psi(t_0, \vec{y})} \gamma^0 \phi(t_0,\vec{y})\: d^3y \:.
\eeq
It follows from current conservation that for any solutions~$\psi, \phi$ of the Dirac equation,
this scalar product is independent of the choice of~$t_0$.
Since the kernel of the fermionic projector is a solution of the Dirac equation,
one is led to evaluating the integral in~\eqref{print} for~$\phi(y)=P(y,z)$
and~$\overline{\psi(y)} = P(x,y)$.
\begin{Lemma} For any~$t \in \R$, there is the distributional relation
\beq \label{spatialnorm}
2 \pi \int_{\R^3} P \big( x, (t, \vec{y}) \big) \:\gamma^0\: P \big( (t, \vec{y}), z \big)\: d^3y
= -P(x,z)\:.
\eeq
\end{Lemma}
\Proof The identity follows by a straightforward computation. First,
\begin{align*}
\int_{\R^3} & P \big( x, (t, \vec{y}) \big) \:\gamma^0\: P \big( (t, \vec{y}), z \big)\: d^3y \\
&= \int_{\R^3} d^3y \int \frac{d^4k}{(2 \pi)^4}\: e^{-i k(x-y)} \int \frac{d^4q}{(2 \pi)^4}\: e^{-i q(y-z)}
\: P_m(k)\:\gamma^0\: P_m(q) \\
&= \int \frac{d^4k}{(2 \pi)^4} \int_\R \frac{d \lambda}{2 \pi}\; e^{-i k x + i q z} 
\: P_m(k)\:\gamma^0\: P_m(q) \Big|_{q = (\lambda, \vec{k})} \:.
\end{align*}
Setting~$k=(\omega, \vec{k})$, we evaluate the $\delta$-distributions inside the factors~$P_m$,
\begin{align*}
\delta(k^2-m^2)& \, \delta(q^2-m^2) \big|_{q = (\lambda, \vec{k})}
= \delta \big( \omega^2 - |\vec{k}|^2 -m^2 \big) \:
\delta \big( \lambda^2 - |\vec{k}|^2 -m^2 \big) \\
&= \delta(\lambda^2 - \omega^2)\: \delta \big( \omega^2 - |\vec{k}|^2 -m^2 \big) \:.
\end{align*}
This shows that we only get a contribution if~$\lambda=\pm \omega$.
Using this fact, we can simplify the Dirac matrices according to
\begin{align*}
(\slashed{k}+m) &\:\gamma^0\: (\slashed{q} + m)
= (\omega \gamma^0 + \vec{k} \vec{\gamma} + m) \,\gamma^0\,
(\pm \omega \gamma^0 + \vec{k} \vec{\gamma} + m) \\
&= (\omega \gamma^0 + \vec{k} \vec{\gamma} + m) \,
(\pm \omega \gamma^0 - \vec{k} \vec{\gamma} + m)\,\gamma^0 \\
&= \Big( (\pm \omega^2 + |\vec{k}|^2 + m^2) \,\gamma^0
+(1 \pm 1) \,\omega\,(\vec{k} \vec{\gamma}) + (1 \pm 1)\, m \omega \Big) \\
&= \left\{ \begin{array}{cl}
2 \omega\, (\slashed{k}+m) & \text{in case~$+$} \\
0 & \text{in case~$-\:.$} \end{array} \right.
\end{align*}
Hence we only get a contribution if~$\lambda=\omega$, giving rise to the identity
\[  \delta(\lambda^2 - \omega^2) = \frac{1}{2 |\omega|}\: \delta(\lambda-\omega)\:. \]
Putting these formulas together, we obtain
\begin{align*}
\int_{\R^3} & P \big( x, (t, \vec{y}) \big) \:\gamma^0\: P \big( (t, \vec{y}), z \big)\: d^3y \\
&= \int \frac{d^4k}{(2 \pi)^4} \int_\R \frac{d \lambda}{2 \pi}\; e^{-i k (x-z)} 
\:\delta(\lambda - \omega)\: \delta( k^2 -m^2)\:
\frac{2 \omega}{2 |\omega|}\, (\slashed{k}+m)\: \Theta(-k^0) \\
&= -\frac{1}{2 \pi} \int \frac{d^4k}{(2 \pi)^4} \; e^{-i k (x-z)} 
\: \delta( k^2 -m^2)\: (\slashed{k}+m)\: \Theta(-k^0) \:.
\end{align*}
This gives the result.
\QED
We refer to~\eqref{spatialnorm} as the {\bf{spatial normalization}} of the fermionic projector.

\subsection{General Discussion of the Normalization Method}
We are interested in the Dirac equation with an interaction via an external potential
\beq \label{direx}
(i \Pdd + \B - m) \,\psi(x) = 0 \:.
\eeq
Here~$\B(x)$ is a matrix-valued potential which we assume to be smooth and symmetric with respect
to the spin scalar product, i.e.\ $\overline{\psi} (\B \phi) = (\overline{\B \psi}) \phi$
(where~$\overline{\psi} \equiv \psi^* \gamma^0$ is the adjoint spinor).
Before beginning the detailed analysis, we now explain and discuss in general terms how the
fermionic projector can be defined and normalized for such interacting systems.

We first point out that the distribution~\eqref{Fourier1} is supported on the lower mass shell.
In this way, the fermionic projector of the vacuum~\eqref{Fourier2} involves a decomposition of the
solution space into the subspaces of positive and negative frequency: only the
solutions of negative frequency on the lower mass shell are taken into account.
Thus in order to get a reasonable definition of the fermionic projector
in the presence of an external potential, one needs to extend the decomposition of the solution
space to the Dirac equation~\eqref{direx}.
In the special case that~$\B$ is static, one can still separate the time dependence
by the plane wave ansatz~$\psi(t, \vec{x}) = e^{-i \omega t} \, \psi_\omega(\vec{x})$, so that
the sign of~$\omega$ gives a canonical splitting of the solution space.
In the general time-dependent setting, however, no plane wave ansatz can be used,
and it is therefore no longer obvious if there still is a {\em{canonical decomposition of the solution space
into two subspaces}}.
This question was answered affirmative in~\cite{sea}, and a perturbative construction of the
decomposition was given. A general conclusion from this analysis is that in order to construct the decomposition,
it is not sufficient to consider the solution space at some fixed time~$t$,
but one must analyze the behavior of the solutions {\em{globally in space-time}}.
In technical terms, this leads to operator products as in~\eqref{massnorm} which, if
written in position space, involve the integral over all of space-time,
\[ (P_m \: P_{m^\prime})(x,y) = \int P_m(x,z) \: P_{m^\prime}(z,y) \:d^4z \:. \]
In order to make sense of such operator products, one must necessarily consider the mass as
a variable parameter and use a $\delta$-normalization as in~\eqref{massnorm}.
This is why the mass normalization arises naturally in the construction of the
fermionic projector. It seems the proper procedure from a mathematical perspective if the
functional calculus in the mass parameter is taken seriously.
For this reason, the fermionic projector was first constructed using the mass normalization
in~\cite{sea, grotz}.

It turns out that in the presence of an external potential~\eqref{direx},
the fermionic projector with mass normalization does in general {\em{not}} satisfy the spatial
normalization condition~\eqref{spatialnorm}.
This incompatibility of the normalization methods is a subtle point which
can be seen explicitly by computing the corresponding normalization integrals
in a perturbation expansion to second order in~$\B$ (alternatively, the differences also become
clear by comparing the perturbation expansions with mass normalization and spatial normalization
as listed in Appendix~\ref{appA} below).
The fermionic projector with spatial normalization can be obtained from
the mass-normalized fermionic projector by a rescaling procedure.
To avoid confusion, we point out that, no matter which normalization procedure is used,
the spectral calculus in the mass parameter is needed in any case
in order to obtain the decomposition of the solution space into
two subspaces. Therefore, the $\delta$-normalization in the mass parameter as in~\eqref{massnorm}
cannot be avoided as an intermediate step of the construction, even if the spatial normalization is used.

Since the mass normalization and the spatial normalization are not compatible,
one must decide which of the two normalization methods to use.
There are no compelling physical reasons for working with one or the other
normalization method. 
Instead, it is an open question which normalization method should be used.
Ultimately, this question can be answered only by physical experiments
(for example by considering loop diagrams as worked out in Section~\ref{secloop} below).
Nevertheless, there are a few arguments in favor of the spatial normalization:
\begin{itemize}
\item[(a)] The spatial integral in~\eqref{spatialnorm} is closely related to the
probability integral for Dirac wave functions. More precisely, the
condition~\eqref{spatialnorm} 
can be understood by saying that all the states of the fermionic projector should be normalized
(up to the irrelevant factor of~$2 \pi$) with respect to the integral over the probability density
\beq \label{prdens}
\int_{\R^3} (\overline{\psi} \gamma^0 \psi)(t, \vec{x})\: d^3x\:.
\eeq
Therefore, the spatial normalization condition seems to be adjusted to the probabilistic interpretation
of the Dirac wave function.
\item[(b)] In the framework of causal fermion systems (as introduced in~\cite{rrev}),
the mass normalization is implemented if one chooses the scalar product on the particle space equal
to the space-time inner product
\beq \label{stip}
\bra \psi|\phi \ket = \int_M (\overline{\psi}\phi)(x) \: d^4x\:,
\eeq
restricted to the occupied fermionic states of the system (in order to make sense of the integral
in~\eqref{stip}, it may necessary to introduce an infrared regularization or to 
use again a $\delta$-normalization in the mass parameter similar to~\eqref{massnorm}).
However, this procedure only works if the inner product~\eqref{stip} is negative definite
on the occupied fermionic states. Since it is not clear why this should always be the case,
the mass normalization does not seem compatible with the framework of causal fermion systems.
\item[(c)] If the image of the fermionic projector is negative definite with respect to~\eqref{stip},
one can construct a corresponding causal fermion system (at least after introducing a regularization).
But this leads to the complicated situation that there are two different scalar products: First, the inner
product~$-\bra .|. \ket$ restricted to the image of the fermionic projector (which coincides with the scalar
product~$\la .|. \ra_\H$ on the particle space~$\H$ of the corresponding causal fermion system). Second,
the scalar product~\eqref{print} obtained by polarizing~\eqref{prdens}
which is needed for the probabilistic interpretation.

Working with the spatial normalization, on the other hand,
it suffices to consider only the scalar product~\eqref{print}.
\item[(d)] The mass normalization only makes sense in space-times of infinite life-time.
The spatial normalization, however, can be used on any globally hyperbolic space-time
(for details see~\cite{finite, infinite}).
\end{itemize}
In view of these arguments, the authors consider the spatial normalization as being more natural
and conceptually more convincing. However, the arguments do not seem strong enough for
making a final judgment. For this reason, in what follows we shall consider both normalization
conditions in parallel and on the same footing. This also has the advantage that we
can work out the similarities and differences in detail.

\section{The Causal Perturbation Expansion with Contour Integrals} \label{seccontour}
We now give a convenient method for deducing all the contributions
to the causal perturbation expansion including the combinatorial factors.
The method is to introduce a resolvent and to recover the fermionic projector
as a suitable Cauchy integral.

\subsection{Preliminaries}
In preparation, we fix our notation and recall a few constructions from~\cite{sea, grotz}.
Starting from the plane-wave solutions of the vacuum Dirac equation, the
equation in the external potential~\eqref{direx} can be solved in a perturbation expansion in~$\B$.
In the language of Feynman diagrams, this is an expansion in terms of tree diagrams.
These diagrams are all well-defined and finite, provided that the potential~$\B$ is sufficiently regular
and has suitable decay properties at infinity (for details see for example~\cite[Lemma~2.2.2]{PFP}).
With this in mind, all our perturbation expansions are well-defined on the level of formal power
series in~$\B$. The questions of convergence of the perturbation expansions can be answered
by using non-perturbative constructions (see~\cite{finite, infinite, hadamard}).
Here we shall not consider such convergence questions, but instead we focus on working out
the properties of the resulting expansions. For notational simplicity we always restrict
attention to one family of Dirac particles of mass~$m$. The generalization to systems of
several families or generations of particles is straightforward using the methods
in~\cite[\S5.1]{PFP} and~\cite[Section~4]{sector}.

We always denote the objects in the presence of the external field by a tilde.
The solutions of the vacuum Dirac equation on the upper respectively lower mass cone are
described by the distributions
\beq \label{Ppmdef}
P_\pm = \frac{1}{2} \,\big(p_m \pm k_m) \:,
\eeq
where
\begin{align}
p_m(q) &= (\slashed{q}+m)\: \delta(q^2-m^2) \label{pdef} \\
k_m(q) &= (\slashed{q}+m)\: \delta(q^2-m^2)\: \epsilon(q^0) \:, \label{kdef}
\end{align}
where~$\epsilon$ denotes the step function~$\epsilon(x)=1$ if~$x>0$ and~$\epsilon(x)=-1$ otherwise.
Moreover, we denote the advanced and retarded Green's functions by
\beq \label{sadret}
s^\vee_m(q) =\lim_{\varepsilon \searrow 0}\: \frac{\slashed{q}+m}{q^2-m^2-i \varepsilon q_0} \qquad \text{and} \qquad
s^\wedge_m(q) =\lim_{\varepsilon \searrow 0}\: \frac{\slashed{q}+m}{q^2-m^2+i \varepsilon q_0} \:.
\eeq
Using the formula
\[ \lim_{\varepsilon \searrow 0} \left( \frac{1}{x-i\varepsilon}
- \frac{1}{x+i \varepsilon} \right) \;=\; 2 \pi i \: \delta(x) \:, \]
one immediately verifies that the distribution~$k_m$ can be expressed in terms of these
Green's functions by
\beq \label{ksrel}
k_m = \frac{1}{2 \pi i}\: (s^\vee_m - s^\wedge_m) \:.
\eeq
In particular, the distribution~$k_m$ is causal in the sense that it vanishes identically
for spacelike separated points.
Moreover, the symmetric Green's function~$s_m$ is defined by
\beq \label{sdef}
s_m = \frac{1}{2}\: (s^\vee_m + s^\wedge_m) \:.
\eeq

In the presence of an external potential~$\B$, the perturbation expansion for
the advanced and retarded Green's functions is unique by causality,
\beq \tilde{s}_m^{\vee}=\sum_{n=0}^{\infty}(-s_m^{\vee}\mathscr{B})^ns_m^{\vee}\;, \qquad
\tilde{s}_m^{\wedge}=\sum_{n=0}^{\infty}(-s_m^{\wedge}\mathscr{B})^ns_m^{\wedge}\:.
\label{series-scaustilde}
\eeq
Using~\eqref{ksrel}, we also have a unique perturbation expansion for the
causal fundamental solution,
\beq
\tilde{k}_m = \frac{1}{2\pi i}(\tilde{s}_m^{\vee}-\tilde{s}_m^{\wedge}) \:. \label{def-ktil}
\eeq
Using the identities
\beq \label{scaus}
s_m^{\vee} = s_m+i\pi k_m\:, \qquad s_m^{\wedge} = s_m-i\pi k_m \:,
\eeq
one can write the above perturbation series as operator product expansions. More precisely,
\beq \label{kex}
\tilde{k}_m=\sum_{\beta=0}^{\infty}(i\pi)^{2\beta} \:b_m^<\, k_m\, (b_m k_m)^{2\beta}\, b_m^>\:,
\eeq
where the factors~$b_m^\bullet$ are defined by
\beq \label{bmdef}
b_m^<=\sum_{n=0}^{\infty}(-s_m\mathscr{B})^n\;,\qquad
b_m=\sum_{n=0}^{\infty}(-\mathscr{B}s_m)^n\mathscr{B}\;,\qquad
b_m^>=\sum_{n=0}^{\infty}(-\mathscr{B}s_m)^n\:.
\eeq

In the following constructions, we need to multiply the operator products in~\eqref{kex}.
These products have a mathematical meaning as distributions in the involved mass parameters,
\begin{align}
p_m\,p_{m'}&=k_m\,k_{m'}=\delta(m-m')\:p_m \\
p_m\,k_{m'}&=k_m\,p_{m'}=\delta(m-m')\:k_m \\
k_m \,b_m^>\, b_{m'}^<\, k_{m'} &= \delta(m-m') \Big( p_m
+ \pi^2\,k_m\, b_m\, p_m\, b_m\, k_m \Big) \:. \label{bbprod}
\end{align}
Since these formulas all involve a common prefactors~$\delta(m-m')$,
we can introduce a convenient notation by leaving out this factor and omitting the
mass indices. For clarity, we denote this short notation with a dot, i.e.\ symbolically
\beq \label{sdotdef}
A \sdot B = C \qquad \text{stands for} \qquad
A_m \,B_{m'} = \delta(m-m')\: C_m \:.
\eeq
With this short notation, the multiplication rules can be written in the compact form
\beq p \sdot p =k \sdot k =p\:, \qquad p \sdot k=k \sdot p=k\:,\qquad
k \,b^> \sdot b^< k =p+\pi^2 kbpbk \:. \label{rules}
\eeq
In all the subsequent calculations, the operator products are well-defined
provided that the potential~$\B$ is sufficiently smooth and has suitable decay properties
at infinity (for details see~\cite[Lemma~2.2.2]{PFP}). However, all infinite series are to be understood
merely as formal power series in the potential~$\B$.

\subsection{The Fermionic Projector with Mass Normalization}
Writing~\eqref{kex} as
\beq \label{ktildef}
\tilde{k} =\sum_{\beta=0}^{\infty}(i\pi)^{2\beta} \:b^<\, k\, (b k)^{2\beta}\, b^>\:,
\eeq
powers of the operator~$\tilde{k}$ with the product~\eqref{sdotdef}
are well-defined using the multiplication rules~\eqref{rules}.
This makes it possible to develop a spectral calculus for~$\tilde{k}$.
In particular, in~\cite{grotz} the operator~$P^\sea$ is constructed as the projection operator
on the negative spectral subspace of~$\tilde{k}$.
We now give an equivalent construction using contour integrals, which gives a more systematic
procedure for computing all the contributions to the expansion.

We introduce the resolvent by
\beq \label{Resdef}
\tilde{R}_\lambda = (\tilde{k} - \lambda)^{-1}\:.
\eeq
We choose a contour~$\Gamma_-$ which encloses the point~$-1$ in counter-clockwise 
direction and does not enclose the points~$1$ and~$0$. We set
\beq \label{Pres}
P^\sea_\res = -\frac{1}{2 \pi i} \ointctrclockwise_{\Gamma_-} \tilde{R}_\lambda\: d\lambda \:.
\eeq
This formula is to be understood as an operator product expansion, as we now explain.
We write~$\tilde{k}$ as
\[ \tilde{k} = k + \Delta k \:, \]
where~$k$ is the corresponding distribution in the vacuum. Then~$\tilde{R}_\lambda$
can be computed as a Neumann series,
\beq \label{tR}
\tilde{R}_\lambda = (k - \lambda + \Delta k)^{-1}
= (1 + R_\lambda \sdot \Delta k)^{-1} \sdot R_\lambda 
= \sum_{n=0}^\infty (-R_\lambda \sdot \Delta k)^n \sdot R_\lambda \:.
\eeq
According to~\eqref{rules}, the operator~$k$ has the eigenvalues~$\pm 1$ and~$0$
with corresponding spectral projectors~$(p \pm k)/2$ and~$\1-p$.
Hence we can write the free resolvent as
\[ R_\lambda = \frac{p+k}{2} \left( \frac{1}{1-\lambda} \right) + \frac{p-k}{2} \left( \frac{1}{-1-\lambda} \right)
- \frac{\1-p}{\lambda} \:. \]
Substituting this formula in~\eqref{tR}, to every order in perturbation theory we obtain
a meromorphic function in~$\lambda$ having poles only at~$\lambda=0$ and~$\lambda= \pm1$.
Thus the contour integral~\eqref{Pres} can be computed with residues, and the result
is independent of the choice of~$\Gamma_-$. In this way, we obtain a perturbation expansion
for~$P^\sea_\res$.

\begin{Prp} \label{prpmass}
The perturbation expansion~$P^\text{\rm{sea}}_\res$ has the properties
\begin{align}
(i \Pdd+\B-m)\, P^\text{\rm{sea}}_\res &= 0 \label{Psol} \\
\big( P^\text{\rm{sea}}_\res \big)^* &= P^\text{\rm{sea}}_\res \label{Pstar} \\
P^\text{\rm{sea}}_\res \sdot P^\text{\rm{sea}}_\res &= P^\text{\rm{sea}}_\res\:. \label{massnorm2}
\end{align}
\end{Prp} \noindent
In view of our notation of omitting the factors~$\delta(m-m')$ introduced before~\eqref{rules},
the idempotence relation~\eqref{massnorm2} agrees precisely with the normalization~\eqref{massnorm}.
Therefore, $P^\sea_\res$ is the {\bf{fermionic projector with mass normalization}}.
The notation for the index ``res'' has evolved historically and has a twofold meaning.
It was first introduced in~\cite{light} to denote the operator~$\tilde{p}^\res$ obtained
by applying to~$\tilde{k}$ the so-called {\em{residual argument}}
(see also the proof of Theorem~\ref{thmlight} below). In~\cite{grotz}, the index ``res''
denoted the operators obtained by {\em{rescaling}} the Dirac sea.
Using the same notation with a different meaning was motivated by the fact that 
the residual argument and the rescaling procedure gave rise to very similar operator product
expansions. What was then considered a surprising coincidence 
can in fact be understood systematically by the symmetry consideration in Section~\ref{secsymm}.

In preparation for the proof of Proposition~\ref{prpmass}, we prove a
spectral calculus for contour integrals which generalizes~\eqref{Pres}.
To this end, let~$f$ be a holomorphic function defined on an
an open neighborhood of the points~$\pm 1$. We define~$f(\tilde{k})$ 
by
inserting the function~$f(\lambda)$ into the contour integral~\eqref{Pres} and integrating
around both spectral points~$\pm 1$,
\beq \label{fres}
f \big( \tilde{k} \big) := -\frac{1}{2 \pi i} \ointctrclockwise_{\Gamma_+ \cup \Gamma_-} f(\lambda)\:
\tilde{R}_\lambda\: d\lambda  \:, 
\eeq
where~$\Gamma_+$ is a contour which encloses~$+1$ in counter-clockwise orientation
(and does not enclose~$-1$ or~$0$).
\begin{Thm} {\bf{(functional calculus)}} \label{thmfcalc}
For any functions~$f, g$ which are holomorphic in discs around~$\pm1$
which contain the contours~$\Gamma_\pm$,
\begin{align}
(i \Pdd+\B-m)\, f \big( \tilde{k} \big) &= 0 \label{fsol} \\ 
f \big( \tilde{k} \big)^* &= \overline{f}\big( \tilde{k} \big) \label{fstar} \\ 
f \big( \tilde{k} \big) \sdot g \big( \tilde{k} \big) &= (fg)\big( \tilde{k} \big) \:. \label{fg} 
\end{align}
\end{Thm}
\Proof 
Since the image of the operator~$\tilde{k}$ lies in the kernel of the Dirac operator, we know that
\[ (i \Pdd + \B - m)\: \tilde{R}_\lambda = (i \Pdd + \B - m)\, \big( -\lambda^{-1} \big) . \]
Taking the contour integral~\eqref{fres} gives~\eqref{fsol}.

The operators~$p_m$, $k_m$ and~$s_m$ are obviously symmetric (see~\eqref{pdef}, \eqref{kdef}
and~\eqref{sdef}). According to~\eqref{kex}, the operator~$\tilde{k}_m$ is also symmetric. Hence
the resolvent~$\tilde{R}_\lambda$ defined by~\eqref{Resdef} has the property
\[ \tilde{R}_\lambda^* = \tilde{R}_{\overline{\lambda}}\:. \]
The relation~\eqref{fstar} follows by taking the adjoint of~\eqref{fres} and
reparametrizing the integral.

The starting point for proving~\eqref{fg} is the resolvent identity
\beq \label{rr1}
\tilde{R}_\lambda \sdot \tilde{R}_{\lambda'} = \frac{1}{\lambda-\lambda'} \left(
\tilde{R}_\lambda - \tilde{R}_{\lambda'} \right) \:.
\eeq
We set~$\Gamma=\Gamma_+ \cup \Gamma_-$ and denote the corresponding
contour for~$\lambda'$ by~$\Gamma'$.
Since the integral~\eqref{fres} is independent of the precise choice of the contour,
we may choose
\[ \Gamma = \partial B_{\delta}(1) \cup \partial B_{\delta}(-1) \qquad \text{and} \qquad
\Gamma'=\partial B_{2 \delta}(1) \cup \partial B_{2 \delta}(-1) \]
for sufficiently small~$\delta<1/2$.
Then~$\Gamma$ does not enclose any point of~$\Gamma'$, implying that
\beq \label{rr2}
\ointctrclockwise_\Gamma \frac{f(\lambda)}{\lambda-\lambda'} \: d\lambda = 0 
\qquad \text{for all~$\lambda' \in \Gamma'$}\:.
\eeq
On the other hand, $\Gamma'$ encloses every point of~$\Gamma$, so that
\beq \label{rr3}
\ointctrclockwise_{\Gamma'} f(\lambda)\, g(\lambda')\: \frac{\tilde{R}_{\lambda}}{\lambda-\lambda'} \: d\lambda' =
-2 \pi i\,  f(\lambda)\, g(\lambda)\: \tilde{R}_\lambda \qquad \text{for all~$\lambda \in \Gamma$}\:.
\eeq
Combining~\eqref{rr1} with~\eqref{rr2} and~\eqref{rr3}, we obtain
\begin{align*}
f \big( \tilde{k} \big) \sdot g \big( \tilde{k} \big) 
&= -\frac{1}{4 \pi^2} \ointctrclockwise_{\Gamma} f(\lambda)\, d\lambda \ointctrclockwise_{\Gamma'} g(\lambda')\, d\lambda'\;
\frac{1}{\lambda-\lambda'} \left( \tilde{R}_\lambda - \tilde{R}_{\lambda'} \right) \\
&= -\frac{1}{2 \pi i} \ointctrclockwise_{\Gamma} f(\lambda)\, g(\lambda)\: \tilde{R}_\lambda \:d\lambda
= (fg)\big( \tilde{k} \big) \:. 
\end{align*}
This concludes the proof.
\QED

\Proof[Proof of Proposition~\ref{prpmass}]
Follows immediately from Theorem~\ref{thmfcalc} if we choose the functions~$f$ and~$g$
to be identically zero in a neighborhood of~$+1$ and identically equal to one in a neighborhood of~$-1$.
\QED

\subsection{The Fermionic Projector with Spatial Normalization} \label{secspatial}
We now turn attention to the spatial normalization integral in~\eqref{spatialnorm}.
We define
\beq \label{Psea}
P^\sea = -\frac{1}{2 \pi i} \ointctrclockwise_{\Gamma_-} (-\lambda)\: \tilde{R}_\lambda\: d\lambda \:.
\eeq
with~$\Gamma_-$ and~$\tilde{R}_\lambda$ as in~\eqref{Pres}.
\begin{Prp} \label{prpspatial}
The expansion~$P^\text{\rm{sea}}$ has the properties
\begin{align}
(i \Pdd+\B-m)\, P^\text{\rm{sea}} &= 0 \label{Psol2} \\
2 \pi \int_{\R^3} P^\text{\rm{sea}} \big( x, (t, \vec{y}) \big) \:\gamma^0\: P^\text{\rm{sea}}
\big( (t, \vec{y}), z \big)\: d^3y &= -P^\text{\rm{sea}}(x,z)\:. \label{Pspatial}
\end{align}
\end{Prp}
The remainder of this section is devoted to the proof of this proposition.
For the spatial integral in~\eqref{Pspatial} we introduce the short notation~$|_t$, i.e.
\beq \label{bardef}
(A \,|_t\, B)(x,z) := 2 \pi \int_{\R^3} A \big( x, (t, \vec{y}) \big) \:\gamma^0\: B \big( (t, \vec{y}), z \big)\: d^3y \:.
\eeq
We begin with a preparatory lemma.
\begin{Lemma} \label{lemmakmid}
For any~$t_0 \in \R$, the distribution~\eqref{def-ktil} has the property
\[ \tilde{k}_m \,|_{t_0}\, \tilde{k}_m = \tilde{k}_m \:. \]
\end{Lemma}
\Proof Clearly, it suffices to prove the relation when evaluated by a test function~$f$.
Then~$\tilde{\phi} := \tilde{k}_m(f)$ is a smooth solution of the Dirac equation with
spatially compact support. Therefore, it suffices to show that for any such solution,
\[ \tilde{\phi}(t, \vec{x}) = 2 \pi \int_{\R^3} \tilde{k}_m(t,\vec{x}; t_0, \vec{y})\,
\gamma^0\, \tilde{\phi}_0(\vec{y})\: d^3y\:. \]
Since~$\tilde{\phi}$ and~$\tilde{k}_m$ satisfy the Dirac equation, it suffices to prove
this equation in the case~$t>t_0$. In this case, the equation simplifies in view of~\eqref{def-ktil} to
\[ \tilde{\phi}(x) = i \int_{\R^3} \tilde{s}^\wedge_m(x,y)\, \gamma^0\, \tilde{\phi}_0(y) \big|_{y=(t_0, \vec{y})}
\: d^3y\:, \]
where we set~$x=(t, \vec{x})$.
This identity is derived as follows: We choose a non-negative function~$\eta \in C^\infty(\R)$
with~$\eta|_{[t_0, t]} \equiv 1$ and~$\eta_{(-\infty, t_0-1)} \equiv 0$.
We also consider~$\eta$ as a function on the time variable in space-time. Then
\[ \tilde{\phi}(x) = (\eta \tilde{\phi})(x) = \tilde{s}^\wedge_m \big( (i \Pdd + \B - m) (\eta \tilde{\phi}) \big)
= \tilde{s}^\wedge_m \big( i \gamma^0 \,\dot{\eta}\, \tilde{\phi}) \big) \:, \]
where we used the defining equation of the Green's function~$\tilde{s}_m^\wedge (i \Pdd_x +\B- m)=\1$
together with the fact that~$\tilde{\phi}$ is a solution of the Dirac equation.
To conclude the proof, we choose a sequence~$\eta_l$ such that
the sequence of derivatives~$\dot{\eta}_l$ converges as~$l \rightarrow \infty$ in the
distributional sense to the $\delta$-distribution~$\delta_{t_0}$ supported at~$t_0$. Then
\begin{align*}
\tilde{s}^\wedge_m \big( i \gamma^0 \,\dot{\eta}\, \tilde{\phi}) \big)(x)
&= \int \left( \tilde{s}^\wedge_m(x,y) \big( i \gamma^0 \, \dot{\eta}(y^0)\, \tilde{\phi}(y) \big) \right) d^4y \\
&\rightarrow \int_{\R^3} \left( \tilde{s}^\wedge_m(x,y) \big( i \gamma^0 \tilde{\phi}) \right) 
\big|_{y=(t_0, \vec{y})} \:d^3y \:,
\end{align*}
giving the result.
\QED

\Proof[Proof of Proposition~\ref{prpspatial}] 
The relation~\eqref{Psol2} follows similar as in Proposition~\ref{prpmass}.
In order to prove~\eqref{Pspatial}, we integrate the relations
\[ \tilde{R}_\lambda \sdot (\tilde{k}-\lambda) = \1 = (\tilde{k}-\lambda) \sdot \tilde{R}_\lambda \:, \]
to obtain
\[ \ointctrclockwise_{\Gamma_-} \tilde{R}_\lambda \sdot \tilde{k} \: d\lambda
= \ointctrclockwise_{\Gamma_-} \tilde{R}_\lambda\: \lambda\: d\lambda
= \ointctrclockwise_{\Gamma_-} \tilde{k}\, \tilde{R}_\lambda \: d\lambda \:. \]
As a consequence,
\[ P^\text{\rm{sea}} \,|_t\, P^\text{\rm{sea}}
= -\frac{1}{4 \pi^2} \ointctrclockwise_{\Gamma_-} d\lambda \ointctrclockwise_{\Gamma_-'} d\lambda' \:
\tilde{R}_\lambda \sdot \tilde{k} \,|_t\, \tilde{k} \sdot \tilde{R}_{\lambda'} \:, \]
and applying Lemma~\ref{lemmakmid} for~$t_0=t$ gives
\[ P^\text{\rm{sea}} \,|_t\, P^\text{\rm{sea}}
= -\frac{1}{4 \pi^2} \ointctrclockwise_{\Gamma_-} d\lambda \ointctrclockwise_{\Gamma_-'} d\lambda' \:
\tilde{R}_\lambda \sdot \tilde{k} \sdot \tilde{R}_{\lambda'}
= -\frac{1}{4 \pi^2} \ointctrclockwise_{\Gamma_-} \lambda\, d\lambda \ointctrclockwise_{\Gamma_-'} d\lambda' \:
\tilde{R}_\lambda \sdot \tilde{R}_{\lambda'} \:. \]
Now we can again apply~\eqref{rr1} and~\eqref{rr2} (which remains valid if the integrand involves
an additional factor~$\lambda$) as well as~\eqref{rr3}. We thus obtain
\[ P^\text{\rm{sea}} \,|_t\, P^\text{\rm{sea}} =  -\frac{1}{2 \pi i} \ointctrclockwise_{\Gamma_-} \lambda\,
\tilde{R}_\lambda \:d\lambda = -P^\text{\rm{sea}} \:, \]
concluding the proof.
\QED
The resulting perturbation expansion agrees with the expansion
given in~\cite[Section~3]{sea} (although at that time
the spatial normalization property was not considered).

\subsection{A Symmetry between the Mass and the Spatial Normalizations} \label{secsymm}
We now want to compute the spatial normalization integral~\eqref{bardef}
for general operator products involving~$p_m$, $k_m$ and~$s_m$
(see~\eqref{pdef}, \eqref{kdef} and~\eqref{sdef}).
If both operators in the product map to solutions of the Dirac equation, it follows from
the conservation of the Dirac current that the integral is independent of the time~$t$.
If the operator product involves a Green's function, however, the product will
in general depend on~$t$. For example, the integral
\[ 2 \pi \int_{\R^2} p_m(x, (t,\vec{y})) \,\gamma^0\, s_m((t,\vec{y}), z) \]
depends on whether~$t$ lies to the future or past of the space-time point~$z$.
As a convenient notation, we write~$|_{-\infty}$ if the time~$t$ at which the integral~\eqref{bardef}
is performed lies in the past of~$x$ and~$z$. Similarly, $|_{+\infty}$ denotes the
inner product if the time~$t$ in~\eqref{bardef} lies in the future of both~$x$ and~$z$.
With this notation, we have the following computation rules.
\begin{Lemma} \label{lemmarules} For all~$t \in \R$,
\begin{gather}
k_m\,|_t\, k_m = k_m = p_m \,|_t\, p_m \label{rule1} \\
k_m\,|_t\, p_m = p_m = p_m\,|_t\, k_m \label{rule2} \\
\mp k_m\,|_{\pm \infty} \, s_m = i \pi\, k_m = \pm s_m\,|_{\pm \infty} \, k_m \label{rule3} \\
\mp p_m\,|_{\pm \infty} \, s_m = i \pi\, p_m = \pm s_m\,|_{\pm \infty} \, p_m \label{rule4} \\
s_m\,|_{\pm \infty} \, s_m = \pi^2 \,k_m \:. \label{rule5}
\end{gather}
\end{Lemma}
\Proof The first equation in~\eqref{rule1} coincides with Lemma~\ref{lemmakmid}.
In order to prove the second equation in~\eqref{rule1}, we write
\[ p_m = k_m \,\epsilon \:, \]
where~$\epsilon(p) = \epsilon(p^0)$ is the operator which multiplies the upper and lower
mass shell by~$+1$ and~$-1$, respectively. Then
\[ p_m \,|_t\, p_m = \epsilon \,k_m \,|_t\, k_m\, \epsilon = \epsilon \,k_m\, \epsilon = k_m \:. \]
The relations~\eqref{rule2} follows similarly.

In order to prove the remaining rules~\eqref{rule3}--\eqref{rule5}, one uses~\eqref{scaus} to
rewrite~$s_m$ in terms of~$k_m$ and a causal Green's function.
We then arrange that the causal Green's function vanishes by using that~$t$ lies in the future respectively
past of~$x$ and~$z$. For example,
\begin{gather*}
k_m \,|_{+\infty}\, s_m = k_m \,|_{+\infty}\, (s_m^\vee - i \pi k_m)
=  - i \pi\: k_m \,|_{+\infty}\, k_m = -i \pi \,k_m \\
p_m \,|_{+\infty}\, s_m = \epsilon\, k_m \,|_{+\infty}\, s_m = -i \pi \,\epsilon \,k_m
= -i \pi \,p_m \\
s_m \,|_{+\infty}\, s_m = (s_m^\wedge + i \pi k_m) \,|_{+\infty}\, (s_m^\vee - i \pi k_m)
=  \pi^2\: k_m \,|_{+\infty}\, k_m = \pi^2\,k_m \:.
\end{gather*}
The other relations are derived similarly.
\QED

\begin{Lemma} For all~$t \in \R$,
\beq \label{rule6}
k_m \,b_m^> \,|_t\, b_{m'}^<\, k_m = k_m + \pi^2\,k_m\, b_m\, k_m\, b_m\, k_m \:.
\eeq
\end{Lemma}
\Proof Since the operator product~$b_{m'}^<\, k_m$ is a solution of the Dirac equation
in the external potential~$\B$, it follows from current conservation that the
product on the left of~\eqref{rule6} is independent of~$t$. In particular,
\beq \label{tav}
k_m \,b_m^> \,|_t\, b_{m'}^<\, k_m = \frac{1}{2} \:
k_m \,b_m^> \,\Big( |_{+\infty} + |_{-\infty} \Big)\, b_{m'}^<\, k_m \:.
\eeq
Computing the operator products in this way, the contributions by~\eqref{rule3} and~\eqref{rule4}
drop out. Thus we only get a contribution if the factors~$b_m^>$ and~$b_m^<$
either both contain no factor~$s_m$ or both contain at least one factor~$s_m$.
Using the computation rules~\eqref{rule1} and~\eqref{rule5} gives the result.
\QED

Comparing the computation rules~\eqref{rule1}, \eqref{rule2} and~\eqref{rule6}
for the spatial normalization integrals with the corresponding rules for the
operator products in~\eqref{rules}, one obtains agreement when applying the following replacement rules:
\begin{align}
|_t &\longrightarrow \cdot \label{r1} \\
p_m &\longrightarrow k_m \label{r2} \\
k_m &\longrightarrow p_m \label{r3} \\
s_m &\longrightarrow s_m \label{r4}
\end{align}
(where the dot in~\eqref{r1} again refers to the short notation~\eqref{sdotdef}).
The replacement rules~\eqref{r2}--\eqref{r4} were already used in the so-called residual argument
to introduce the operator~$\tilde{p}^\res_m$ (cf.~\cite[eqs~(3.16) and~(3.17)]{sea}).
We write symbolically
\beq
\tilde{k}_m \longrightarrow \tilde{p}^\res_m \:. \label{r5}
\eeq
Combining the rule~\eqref{r1} with Lemma~\ref{lemmakmid}, one finds that
\[ \tilde{p}^\res \sdot \tilde{p}^\res = \tilde{p}^\res \]
(being a short notation for~$\tilde{p}^\res_m \,\tilde{p}^\res_{m'} = \delta(m-m')\, \tilde{p}^\res_m$).
Thus~$\tilde{p}^\res$ has the correct mass normalization. 
This explains why it coincides with the corresponding operator introduced in~\cite{grotz}
by a rescaling procedure for the Dirac sea. It can be written similar to~\eqref{Pres} as the contour integral
\beq \label{ptmres}
\tilde{p}_m^\res = -\frac{1}{2 \pi i} \ointctrclockwise_{\Gamma_+ \cup \Gamma_-} \tilde{R}_\lambda\: d\lambda \:.
\eeq
Next, we introduce the operator~$\tilde{p}_m$ similar to~\eqref{Psea} by
\beq \label{ptm}
\tilde{p}_m = -\frac{1}{2 \pi i} \left( \ointctrclockwise_{\Gamma_+} - \ointctrclockwise_{\Gamma_-} \right) \lambda\: \tilde{R}_\lambda\: d\lambda \:.
\eeq
Repeating the computation in the proof of Proposition~\ref{prpspatial}, one sees that it satisfies the spatial normalization condition
\[ \tilde{p}_m \,|_t\, \tilde{p}_m = \tilde{k}_m \:. \]
Again applying our replacements rules, we obtain an operator~$\tilde{k}^\res_m$,
\beq \label{r6}
\tilde{p}_m \longrightarrow \tilde{k}^\res_m \:,
\eeq
which satisfies the mass normalization condition
\[ \tilde{k}^\res \sdot \tilde{k}^\res = \tilde{p}^\res \:. \]
It can be written similar to~\eqref{Pres} as the contour integral
\beq \label{ktmres}
\tilde{k}_m^\res = -\frac{1}{2 \pi i} \left( \ointctrclockwise_{\Gamma_+} - \ointctrclockwise_{\Gamma_-} \right)
\tilde{R}_\lambda\: d\lambda \:.
\eeq

Finally, we can write the fermionic projector with mass and spatial normalization as
\beq \label{Ppk}
P^\sea_\res = \frac{1}{2} \,\big( \tilde{p}^\res_m - \tilde{k}^\res_m \big)
\qquad \text{and} \qquad
P^\sea = \frac{1}{2} \,\big( \tilde{p}_m - \tilde{k}_m \big) \:.
\eeq
This shows that our replacement rules also transform these
fermionic projectors into each other; more precisely,
\beq
P^\sea \longrightarrow -P^\sea_\res \:. \label{r7}
\eeq
We have thus found a symmetry in the perturbation expansions with mass and spatial
normalization: If in the operator expansions we exchange all
operators according to the replacement rules~\eqref{r2}--\eqref{r4}, then according to~\eqref{r7}
the fermionic projector with spatial normalization is transformed up to minus the fermionic
projector with mass normalization. This symmetry was already observed in~\cite{grotz},
but without understanding the underlying reason~\eqref{r1}.

\section{The Unitary Perturbation Flow} \label{secflow}

\subsection{The Unitary Perturbation Flow with Mass Normalization}
In~\cite[Section~5]{grotz} it is shown that there exists an operator~$U$ which
transforms the vacuum operators~$p_m$ and~$k_m$ to the corresponding
interacting operators with mass normalization~$\tilde{p}_m^\res$ and~$\tilde{k}_m^\res$.
For a consistency, we now denote this operator by~$U_\res$. Then
\beq \label{Udef}
\tilde{P}^\sea_\res = U_\res \sdot \Big( \frac{p_m - k_m}{2} \Big) \sdot U_\res^* \:.
\eeq
The operator~$U_\res$ maps solutions of the vacuum Dirac equation to
solutions of the Dirac equation in the potential. This mapping is invertible, and it is an
isometry with respect to the indefinite inner product~\eqref{stip}.
For simplicity, we say that~$U_\res$ is {\em{unitary}} with respect to the indefinite inner product~\eqref{stip}.
In applications, one considers a family of potentials~$\B(\tau)$ (in the simplest case
the family~$\B(\tau) = \tau \B_0$ which ``turns on'' the interaction)
and considers the corresponding family of unitary transformations~$U_\res(\tau)$.
Then~$U_\res(\tau)$ defines a one-parameter family of transformations,
the so-called {\em{unitary perturbation flow}}.
We now give a systematic procedure for computing the unitary perturbation flow
to any order in perturbation theory.

\begin{Lemma} The operators~$\tilde{k}_m^\res$ and~$\tilde{p}_m^\res$
defined by~\eqref{ktmres} and~\eqref{ptmres} satisfy the relations
\begin{gather}
(i \Pdd + \B - m) \, \tilde{p}^\res = 0 \label{pDir} \\
(\tilde{p}^\res)^* = \tilde{p}^\res = \tilde{p}^\res \sdot \tilde{p}^\res \label{pproj} \\
(\tilde{k}^\res)^* = \tilde{k}^\res = \tilde{k}^\res \sdot \tilde{p}^\res
= \tilde{p}^\res \sdot \tilde{k}^\res \:. \label{pkrel}
\end{gather}
\end{Lemma}
\Proof Follows immediately from the functional calculus of Theorem~\ref{thmfcalc}.
\QED

Our method for computing~$U_\res$ is to ``turn on the perturbation adiabatically.'' Thus
for a parameter~$\tau \in [0,1]$ we let~$\tilde{p}^\res(\tau)$ be the spectral projector corresponding to
the perturbation operator~$\tau \B$. We define~$U_\res^*(\tau)$ by
\beq \label{Usdef}
U_\res^*(\tau) = \lim_{N \rightarrow \infty} \tilde{p}^\res(0) \sdot \tilde{p}^\res \Big( \frac{\tau}{N} \Big)\:
\cdots\: \tilde{p}^\res\Big( \frac{(N-1)\, \tau}{N} \Big) \sdot \tilde{p}^\res(\tau) \:.
\eeq
Then~$U_\res^*(\tau)$ satisfies the differential equation
\begin{align*}
\frac{d}{d\tau} U_\res^*(\tau) &= \lim_{\varepsilon \searrow 0} \frac{U_\res^*(\tau+\varepsilon) - U_\res^*(\tau)}{\varepsilon} \\
&= \lim_{\varepsilon \searrow 0} U_\res^*(\tau) \sdot \frac{\tilde{p}^\res(\tau+\varepsilon)
- \tilde{p}^\res(\tau)}{\varepsilon}
= U_\res^*(\tau) \sdot \Big( \frac{d}{d\tau} \tilde{p}^\res(\tau) \Big) \:.
\end{align*}
Noting that~$U_\res^*(0)=\tilde{p}_\res(0)$ (as is obvious from~\eqref{Usdef} and~\eqref{pproj}), we can solve this differential equation with an ordered exponential,
\beq \label{Uexp}
U_\res^*(\tau) = \tilde{p}^\res(0) \sdot \Pexp \left( \int_0^\tau  (\tilde{p}^\res)'(s) \: ds \right) ,
\eeq
so that
\begin{align}
U_\res^*(\tau) &= \tilde{p}^\res(0) + \tilde{p}^\res(0) \sdot \int_0^\tau (\tilde{p}^\res)'(s)\: ds \notag \\
&\qquad + \tilde{p}^\res(0) \sdot \int_0^\tau ds_1 \int_0^{s_1} ds_2\:
(\tilde{p}^\res)'(s_2) \sdot \,(\tilde{p}^\res)'(s_1)  + \cdots \label{Uresex} \\
&= \tilde{p}^\res(0) \sdot \tilde{p}^\res(\tau) +
\tilde{p}^\res(0) \sdot \int_0^\tau ds_1 \: \big(\tilde{p}^\res(s_1) - \tilde{p}^\res(0) \big)
\sdot \,(\tilde{p}^\res)'(s_1)  + \cdots\:. \notag
\end{align}

We now verify that the resulting operator~$U_\res$ has the required properties.
\begin{Prp} \label{prpflow}
The one-parameter family of operators defined by~\eqref{Usdef} satisfy the Dirac equation and are unitary,
\begin{gather}
(i \Pdd + \tau \B - m)\, U_\res(\tau) = 0 \label{UDir} \\
U(\tau) \sdot U^*(\tau) = \1 = U^*(\tau) \sdot U(\tau)\:. \label{Uunit}
\end{gather}
Moreover, they map the free fundamental solutions and spectral projectors
to the corresponding interacting objects,
\beq
U_\res(\tau) \sdot k \sdot U_\res^*(\tau) = \tilde{k}^\res(\tau) \:,\qquad
U_\res(\tau) \sdot p \sdot U_\res^*(\tau) = \tilde{p}^\res(\tau) \:. \label{Utrans}
\eeq
\end{Prp}
\Proof The Dirac equation~\eqref{UDir} is obviously satisfied in view of~\eqref{pDir} and~\eqref{pproj}
as well as the fact that the operator~$U_\res^*(\tau)$ in~\eqref{Usdef} involves a factor~$\tilde{p}^\res(\tau)$
at the very right.
In order to show unitarity, it suffices to prove the second equality in~\eqref{Uunit}.
Differentiating the first equation in~\eqref{pproj}, we know that
\[ (\tilde{p}^\res)'(\tau) = (\tilde{p}^\res)'(\tau)^*\:, \]
so that we can omit the stars of~$\tilde{p}^\res$ and its derivatives in all calculations.
Next, differentiating the last relation in~\eqref{pproj} gives
\[ (\tilde{p}^\res)'(\tau) \sdot \tilde{p}^\res(\tau) + \tilde{p}^\res(\tau) \sdot\, (\tilde{p}^\res)'(\tau)
= (\tilde{p}^\res)'(\tau) \:. \]
Multiplying from the left and right by~$\tilde{p}^\res$ and using~\eqref{pproj}, we obtain the identity
\[ \tilde{p}^\res(\tau) \sdot (\tilde{p}^\res)'(\tau) \sdot \tilde{p}^\res(\tau) = 0 \:. \]
Since the operator~$U_\res^*(\tau)$ involves a factor~$\tilde{p}^\res(\tau)$ at the right, it follows that
\[ U_\res^*(\tau) \sdot (\tilde{p}^\res)'(\tau) \sdot U_\res(\tau) = 0 \:. \]
Thus
\beq \label{UsUdiff}
\frac{d}{d\tau} \Big( U_\res^*(\tau) \sdot U_\res(\tau) \Big) = 2 \:U_\res^*(\tau) \sdot (\tilde{p}^\res)'(\tau) \sdot U_\res(\tau) = 0 \:.
\eeq
For~$\tau=0$, it follows from~\eqref{Usdef} that
\beq \label{UsUinitial}
U_\res^*(0) \sdot U_\res(0) = \tilde{p}^\res(0) \sdot \tilde{p}^\res(0) = p \sdot p = p \:,
\eeq
where in the last step we used the calculation rules~\eqref{rules}.
These rules also show that~$p$ acts on the free solutions as the identity. Therefore, we can
also write~\eqref{UsUinitial} as~$U^*(0) \sdot U(0)=\1$.
Integrating~\eqref{UsUdiff} gives the unitarity~\eqref{Uunit}.

The first equation in~\eqref{Utrans} follows similarly from the fact that
\begin{align*}
U_\res^*(0) \sdot \tilde{p}^\res(0) \sdot U_\res(0) &=\1 \qquad \text{and} \\
\frac{d}{d\tau} \Big( U_\res^*(\tau)\,\tilde{p}^\res(\tau)\, U_\res(\tau) \Big)
&= 3 \:U_\res^*(\tau) \sdot (\tilde{p}^\res)'(\tau) \sdot U_\res(\tau) = 0 \:.
\end{align*}
In order to derive the second equation in~\eqref{Utrans}, we differentiate~\eqref{pkrel} to obtain
\begin{align*}
(\tilde{k}^\res)'(\tau) &\sdot \tilde{p}^\res(\tau) + \tilde{k}^\res(\tau) \sdot \,(\tilde{p}^\res)'(\tau)
= (\tilde{k}^\res)'(\tau) \\
&= (\tilde{p}^\res)'(\tau) \sdot \tilde{k}^\res(\tau)
+ \tilde{p}^\res(\tau) \sdot \,(\tilde{k}^\res)'(\tau) \:.
\end{align*}
Multiplying from the left and right by~$\tilde{p}^\res$, we can apply~\eqref{pproj} and~\eqref{pkrel} to get
\[ \tilde{k}^\res(\tau) \sdot (\tilde{p}^\res)'(\tau) \sdot \tilde{p}^\res(\tau) = 0 =
\tilde{p}^\res(\tau) \sdot (\tilde{p}^\res)'(\tau) \sdot \tilde{k}^\res(\tau) = 0 \:. \]
As a consequence,
\begin{align*}
&\frac{d}{d\tau} \Big( U_\res^*(\tau) \sdot \tilde{k}^\res(\tau) \sdot U_\res(\tau) \Big) \\
&\quad = U_\res^*(\tau) \sdot \Big( (\tilde{p}^\res)'(\tau) \sdot \tilde{k}^\res(\tau) \sdot \tilde{p}^\res(\tau)
+ \tilde{p}^\res(\tau) \sdot \tilde{k}^\res(\tau) \sdot (\tilde{p}^\res)'(\tau) \Big) \sdot U_\res(\tau) = 0 \:.
\end{align*}
Using that~$U_\res^*(0) \sdot \tilde{k}^\res(0) \sdot U_\res(0) =k$, the result follows.
\QED

\subsection{The Unitary Perturbation Flow with Spatial Normalization}
We now want to construct an operator~$V$ which introduces the interaction in the
case of a spatial normalization, i.e.\ in analogy to~\eqref{Udef}
\[ \tilde{P}^\sea = U \,|_t\, \Big( \frac{p_m - k_m}{2} \Big) \,|_t\, U^* \:, \]
where the adjoint again refers to the indefinite inner product~\eqref{stip}.
Since the fermionic projector with spatial normalization will in general violate the
mass normalization condition (i.e.\ in general $\tilde{P}^\sea \tilde{P}^\sea \neq
\tilde{P}^\sea$), the operator~$U$ will in general {\em{not}} be unitary
with respect to~\eqref{stip}. But it is unitary with respect to the scalar product~\eqref{print},
in the following sense: The scalar product~\eqref{print} is time independent
on the solution space of the Dirac equation. The operator~$U$ maps solutions of the vacuum Dirac equation
to the solutions of the Dirac equation in the external potential. Therefore, we have two different solution spaces,
and the scalar product~\eqref{print} on these spaces should be considered as two separate objects.
By unitarity of~$U$ we mean that~$U$ is an isometric bijection of the solutions of the vacuum Dirac equation
to the solutions of the Dirac equation in the external potential.
The simplest way to construct~$U$ is to use the symmetry between the mass and
spatial normalization of Section~\ref{secsymm}. Applying it to Proposition~\ref{prpflow}
gives the following result.

\begin{Prp} \label{prpflowspatial}
The operators~$U(\tau)$ obtained from the operator~$U^\res(\tau)$
by the replacement rules~\eqref{r2}--\eqref{r4} satisfy the Dirac equation
\[ (i \Pdd + \tau \B - m)\, U(\tau) = 0 \:. \]
Moreover, they map the free fundamental solutions and spectral projectors
to the corresponding interacting objects with spatial normalization,
\[ U(\tau) \,|_t\, k \,|_t\, U^*(\tau) = \tilde{k}(\tau) \:,\qquad
U(\tau) \,|_t\, p \,|_t\, U^*(\tau) = \tilde{p}(\tau) \:. \]
The operators~$U(\tau)$ are unitary with respect to the scalar product~\eqref{print},
meaning that
\[ U(\tau) \,|_t\, U(\tau)^* = \1 = U(\tau)^* \,|_t\, U(\tau) \]
(where the star always denotes the adjoint with respect to the inner product~\eqref{stip}).
\end{Prp}

\subsection{Geometric Phases}
We finally note that the operator~$U_\res[\B] := U_\res(1)$ is not uniquely determined by
its properties~\eqref{UDir}--\eqref{Utrans}. In particular, for a given potential~$\B$,
we could have chosen more generally an arbitrary curve~$\B(\tau)$ with~$0 \leq \tau \leq 1$
in the space of all smooth potentials
with~$\B(0)=0$ and~$\B(1)=\B$ and could have replaced the definition~\eqref{Uexp} by
\beq \label{Uphase}
U_\res^* := \tilde{p}^\res(0) \sdot \Pexp \left( \int_0^1  \partial_s \big( \tilde{p}^\res[\B(s)] \big) \, ds \right) .
\eeq
This alternative definition of~$U_\res$ also has all the desired properties.
However, it does depend on the choice of the curve~$\B(\tau)$.
More precisely, the unitary transformations~$U_\res^*$ corresponding to two different
loops may differ by a unitary operator acting on the solution space of the Dirac equation.
This unitary operator can be interpreted as a ``generalized phase transformation.''
If we connect two different curves having the same endpoints, we obtain a closed loop.
Thus the the dependence on the curve~$\B(\tau)$ can be restated by saying that the
unitary perturbation flow involves a generalized geometric phase
which is picked up when the system is changed adiabatically around a closed loop.
In more mathematical terms, one can regard the unitary perturbation flow as
a parallel transport along the curve~$\B(\tau)$. Then the geometric phase
is the holonomy of this parallel transport.

We now explain the similarities and differences of this geometric phase to the well-known Berry
phase~\cite{berry}. We first recall that in the description of the Berry phase,
one changes the potential in a Schr\"odinger operator adiabatically and continually projects onto a
specific bound state (which clearly also varies adiabatically). Similarly, in~\eqref{Usdef} the
potential~$\B(\tau)$ is varied adiabatically, and we obtain a generalized phase.
An obvious difference is that in our setting
the potential is given in space-time, and~$\tau$ parametrizes a family of space-times with different potentials.
As a consequence, it is not clear how the geometric phase in~\eqref{Uphase} could be detected
in an experiment. Namely, in order to see the phase in~\eqref{Uphase} one would have to
consider a family of space-times which change adiabatically with a parameter~$\tau$.
It is unclear to us how the the resulting family of space-times can be reconciled with the fact that an
observer necessarily lives in a fixed space-time. For this reason, we consider the geometric phase
in~\eqref{Uphase} to be more of mathematical than of physical significance.

From the mathematical point of view, an important difference between~\eqref{Uphase}
and the usual Berry phase is that in~\eqref{Usdef} we do not project continually onto a bound state,
but onto the whole solution space of the Dirac equation.
As a consequence, our holonomy is not only a phase of a bound state,
but it is a unitary endomorphism of the solution space of the Dirac equation.
In view of~\eqref{Utrans}, this endomorphism also respects the decomposition
into generalized positive and negative energy solutions.

In order to illustrate the holonomy, we finally consider the simplest possible example.
Let~$\B(\tau)$ a closed loop with~$\B(0)=\B(1)=0$. Then it is shown in Appendix~\ref{appA}
that in second order perturbation theory,
\beq \label{Ures2}
U_\res^*(1) = p + \pi^2 \int_0^1 p \,\B'(s)\, p \,\B(s)\, p \:ds + \O(\B^3) \:.
\eeq
The integral does not vanish along general loops, giving a non-trivial holonomy.
Note that~$U^\res(1)$ maps the solution space of the vacuum Dirac equation
to itself. The integral in~\eqref{Ures2} is anti-symmetric because
\begin{align*}
\Big( & \int_0^1 p \,\B'(s)\, p \,\B(s)\, p \:ds \Big)^* =
\int_0^1 p \,\B(s)\, p \,\B'(s)\, p \:ds \\
&= p \,\B(s)\, p \,\B(s)\, p \Big|_{s=0}^{s=1} - \int_0^1 p \,\B'(s)\, p \,\B(s)\, p \:ds
= - \int_0^1 p \,\B'(s)\, p \,\B(s)\, p \:ds \:,
\end{align*}
which means that~$U_\res^*(1)$ is indeed unitary to second order in perturbation theory.
One can verify by explicit computation that~$U_\res^*(1)$ is also unitary to higher order.

\section{Other Perturbation Expansions of the Fermionic Projector} \label{secother}
As an alternative to the causal perturbation expansion, one can also consider a
retarded expansion in which the potential~$\B$ at a space-time point~$x$
influences the fermionic wave functions only in the causal future of~$x$.
Such a retarded perturbation expansion is physically questionable because it distinguishes
a direction of time. Nevertheless, it is useful in certain applications when the system
(including all the sea states) is in a fixed configuration in the past.
Another possible method is to perform the perturbation expansion exclusively with
the Feynman propagator. This method is again physically questionable, this time because
it works with the notion of positive and negative frequency which 
in curved space-time has no observer-independent meaning.

In this section we work out these alternative perturbation expansions from a mathematical point
of view and collect some of their properties. This is instructive in comparison with the
causal expansion with mass or spatial normalization.

\subsection{The Retarded Perturbation Expansion}
For a Dirac wave function~$\psi$, the retarded perturbation expansion is 
obtained similar to~\eqref{series-scaustilde} by iteratively applying the retarded Green's function, i.e.
\[ \tilde{\psi} = \sum_{n=0}^\infty (-s^\wedge_m \,\B)^n \,\psi \:. \]
In order for our notation to harmonize with that for the perturbation flow, we write
\[ \tilde{\psi} = U_\ret \sdot \psi \qquad \text{with} \qquad U_\ret = \sum_{n=0}^\infty 
(-s^\wedge_m \,\B)^n \,p_m \:. \]
Thinking of the fermionic projector as being composed of bra and ket states, its perturbation expansion
is given similar to~\eqref{Udef} by
\[ \tilde{P}^\sea_\ret = U_\ret \sdot \Big( \frac{p_m - k_m}{2} \Big) \sdot U_\ret^* \:, \]
where the adjoint~$U_\ret^*$ (taken with respect to the indefinite inner product~\eqref{stip}) involves
the advanced Green's function,
\[ U_\ret^* = \sum_{n=0}^\infty p_m\, (-\B s^\vee_m)^n \:. \]

\begin{Prp} \label{prpret} The retarded perturbation expansion of the fermionic projector~$\tilde{P}^\sea_\ret$
has the representation
\beq \label{Pretdef}
\tilde{P}^\sea_\ret = \frac{1}{2} \,\big( \tilde{p}^\ret_m - \tilde{k}_m \big)
\eeq
with~$\tilde{k}_m$ according to~\eqref{def-ktil} and
\[ \tilde{p}_m^\ret := U_\ret \sdot p_m \sdot U_\ret^* \:. \]
The spatial normalization condition is satisfied; i.e., using the notation~\eqref{bardef},
\beq \label{retnorm}
\tilde{P}^\sea_\ret \,|_t\, \tilde{P}^\sea_\ret = \tilde{P}^\sea_\ret 
\qquad \text{for all~$t \in \R$}\:.
\eeq
\end{Prp}
\Proof From~\eqref{ksrel} we have
\begin{align*}
U_\ret \sdot k_m \sdot U_\ret^* &= \frac{1}{2 \pi i} \sum_{n,n'=0}^\infty (-s^\wedge_m \,\B)^n
\Big( s_m^\vee - s_m^\wedge \Big) (-\B s^\vee_m)^{n'} \:. \\
\intertext{Using that the sums are telescopic, we obtain}
U_\ret \sdot k_m \sdot U_\ret^* &= \frac{1}{2 \pi i}
\sum_{n'=0}^\infty s_m^\vee (-\B s^\vee_m)^{n'} - \sum_{n=0}^\infty (-s^\wedge_m \,\B)^n s_m^\wedge \\
&=  \frac{1}{2 \pi i} \big( \tilde{s}^\vee_m - \tilde{s}^\wedge_m \big) = \tilde{k}_m \:,
\end{align*}
where in the last line we used~\eqref{series-scaustilde} and~\eqref{def-ktil}.
Hence for the operator~$k_m$, the retarded perturbation expansion coincides with the
causal perturbation expansion. This proves~\eqref{Pretdef}.

The spatial normalization condition can be verified in two different ways.
The first method uses the fact that, again due to current conservation, it suffices to prove~\eqref{retnorm}
for any~$t$. In the limit~$t \rightarrow -\infty$, $P^\sea_\ret$ goes over to the vacuum fermionic
projector, so that we can use~\eqref{spatialnorm}.
The second method is to verify the spatial normalization condition directly using the
computation rules of Lemma~\ref{lemmarules}. Since~$P^\sea_\ret$ satisfies the Dirac equation,
exactly as in~\eqref{tav} we may take the mean of the computation rules at~$t=\pm \infty$.
We use the short notation
\[ | = \frac{1}{2} \,\big( |_{+\infty} + |_{-\infty} \big) . \]
Decomposing the Green's functions
according to~\eqref{scaus}, we obtain the computation rules
\begin{align*}
(p_m-k_m) \,|\, s_m^\wedge &=  (p_m-k_m) \,|\, (s_m-i\pi k_m) = -i \pi \, (p_m-k_m) \\
s_m^\vee \,|\, (p_m-k_m) &= (s_m+i\pi k_m) \,|\, (p_m-k_m) = i \pi \, (p_m-k_m) \\
s_m^\vee \,|\, s_m^\wedge &= (s_m+i\pi k_m) \,|\,(s_m-i\pi k_m) = 2 \pi^2 \,k_m \:.
\end{align*}
Hence
\begin{align}
4\, \big( &\tilde{P}^\sea_\ret \,|\, \tilde{P}^\sea_\ret - \tilde{P}^\sea_\ret \big)
= U_\ret \sdot \big( p_m - k_m \big) \sdot U_\ret^* \,|\, U_\ret \sdot \big( p_m - k_m \big) \sdot U_\ret^*
- 4 \tilde{P}^\sea_\ret \nonumber \\
=\,& \sum_{n=1}^\infty U_\ret \sdot \big( p_m - k_m \big)\,|\, (-s^\wedge_m \,\B)^n \:\big( p_m - k_m \big)
\sdot U_\ret^* \nonumber \\
&+ \sum_{n'=1}^\infty
U_\ret \sdot \big( p_m - k_m \big)\: (-\B\, s^\vee_m)^{n'} \,|\, \big( p_m - k_m \big) \sdot U_\ret^* \nonumber \\
&+ \sum_{n,n'=1}^\infty
U_\ret \sdot \big( p_m - k_m \big)\: (-\B\, s^\vee_m)^{n'} \,|\, 
(-s^\wedge_m \,\B)^n \big( p_m - k_m \big) \sdot U_\ret^* \nonumber \\
=\,&-i \pi \sum_{n=0}^\infty U_\ret \sdot \big( p_m - k_m \big) \Big[ \B\, (-s^\wedge_m \,\B)^n
-(-\B\, s^\vee_m)^n \,\B \Big] \big( p_m - k_m \big) \sdot U_\ret^* \label{sum1} \\
&+2 \pi^2 \sum_{n,n'=0}^\infty
U_\ret \sdot \big( p_m - k_m \big)\: (-\B\, s^\vee_m)^{n'} \,\B\,k_m\, \B\,
(-s^\wedge_m \,\B)^n \big( p_m - k_m \big) \sdot U_\ret^* \:. \label{sum2}
\end{align}
The differences of the series involving the advanced and retarded Green's can be
rewritten with the help of~\eqref{ksrel} as a telescopic sum,
\[ \sum_{n=0}^\infty \Big( (-s^\vee_m \,\B)^n - (-s^\wedge_m \,\B)^n \Big) = 
2 \pi i \sum_{n,n'=0}^\infty (-s^\vee_m \,\B)^n (-k_m\, \B) (-s^\wedge_m \,\B)^{n'} \:. \]
Using this relation, the summands in~\eqref{sum1} and~\eqref{sum2} all cancel, giving the result.
\QED

\subsection{The Expansion with Feynman Propagators}
We finally remark that, similar to the retarded Green's function in the
retarded perturbation expansion, one can also perform the perturbation expansion with
any other Green's function. As an example, we consider the perturbation expansion with
the {\em{Feynman propagator}}~$s^+_m$, where
\beq \label{spmdef}
s_m^\pm(k) := \lim_{\varepsilon \searrow 0} \frac{\slashed{k}+m}{k^2 - m^2 \mp i \varepsilon} \:.
\eeq
Then
\beq \label{Updef}
\tilde{\psi} = U_- \cdot \psi \qquad \text{with} \qquad U_- = \sum_{n=0}^\infty (-s^-_m \,\B)^n \, p_m \:.
\eeq
Consequently,
\[ \tilde{P}^\sea_\text{\rm{F}} := U_- \cdot \Big( \frac{p_m - k_m}{2} \Big) \cdot U_-^* \:, \]
where
\[ U_-^* = \sum_{n=0}^\infty p_m\, (-\B s^+_m)^n \:. \]

\begin{Prp} \label{pfpF} The perturbation expansion with Feynman propagators of the fer\-mio\-nic
projector~$P^\sea_\text{\rm{F}}$ has the representation
\beq \label{PFdef}
\tilde{P}^\sea_\text{\rm{F}} = \frac{1}{2} \,\big( \tilde{p}^\res_m - \tilde{k}^\text{\rm{F}}_m \big)
\eeq
with~$\tilde{p}^\res_m$ according to~\eqref{ptmres} and
\beq \label{kfmdef}
\tilde{k}_m^\text{\rm{F}} := U_- \cdot k_m \cdot U_-^* \:.
\eeq
Moreover, $\tilde{P}^\sea_\text{\rm{F}}$ satisfies the mass normalization condition, i.e.\
\[ \tilde{P}^\sea_\text{\rm{F}} \cdot \tilde{P}^\sea_\text{\rm{F}} = \tilde{P}^\sea_\text{\rm{F}} \:. \]
\end{Prp}
\Proof Using the relations
\beq \label{spmdec}
s_m^\pm = s_m \pm i \pi\, p_m \:,
\eeq
we have
\[ U_- \cdot p_m \cdot U_-^* = \frac{1}{2 \pi i} \sum_{n,n'=0}^\infty (-s^-_m \,\B)^n
\Big( s_m^+ - s_m^- \Big) (-\B s^+_m)^{n'} \:. \]
Using that the sums are telescopic, we obtain
\[ U_+ \cdot p_m \cdot U_+^* = \frac{1}{2 \pi i}
\sum_{n'=0}^\infty s_m^+ (-\B s^+_m)^{n'} - \sum_{n=0}^\infty (-s^-_m \,\B)^n s_m^-
= \frac{1}{2 \pi i} \big( \tilde{s}^+_m - \tilde{s}^-_m \big) = \tilde{p}^\res_m \:, \]
where in the last step we used~\eqref{series-scaustilde} and~\eqref{def-ktil}
together with the replacement rules~\eqref{r2}--\eqref{r5}.
This proves~\eqref{PFdef}.

The mass normalization condition can be proved in two ways.
One method is to verify it by explicit computation very similar as in the proof of Proposition~\ref{prpret}
by using~\eqref{spmdec} together with the multiplication rules~\eqref{rules}.
Alternatively, one may relate the mass normalization of~$P^\sea_{\text{\rm{F}}}$ directly
to the statement of Proposition~\ref{prpret} by using the symmetry between the mass and
the spatial normalizations shown in Section~\ref{secsymm}. Namely, 
comparing~\eqref{scaus} with~\eqref{spmdec}, one sees that the rules~\eqref{r1}--\eqref{r4}
give rise to the replacement
\beq \label{retf}
P^\sea_\ret \longrightarrow -P^\sea_{\text{\rm{F}}} \:.
\eeq
Hence the spatial normalization of~$P^\sea_\ret$ corresponds to the mass normalization
of~$P^\sea_\text{\rm{F}}$.
\QED
Combining~\eqref{Pretdef}~\eqref{PFdef}
with the replacement rules~\eqref{r5}~\eqref{retf}, one obtains
\begin{align} \label{r8}
p^\ret_m &\longrightarrow k^\text{\rm{F}}_m \:.
\end{align}

\section{Causality of the Light-Cone Expansion} \label{seclight}
We now work out a few properties of the fermionic projector in position space.
We generalize concepts introduced in~\cite{light} and compare the results for the
different perturbation expansions.

In general terms, each perturbation expansion expresses the fermionic projector as a
sum of operator products of the form
\[ P^\text{sea} = \sum_{k=0}^\infty \sum_{\alpha=0}^{\alpha_{\max}(k)} c_\alpha\;
C_{1,\alpha} \, \B\,C_{2,\alpha} \,\B\, \cdots \,\B\, C_{k+1, \alpha} \:, \]
where the factors~$C_{l,\alpha}$ are the Green's functions~$s_m$ or fundamental solutions~$p_m$,
$k_m$ of the free Dirac equation, and the~$c_\alpha$ are combinatorial factors.
Provided that the potential~$\B$ is smooth and has suitable decay properties at infinity,
any such operator product is a well-defined tempered distribution on~$M \times M$
(for details see~\cite[Lemma~1.1]{light} or~\cite[Lemma~2.2.2]{PFP}).
For the following analysis, it is preferable to express~$p_m$ and~$k_m$ in terms of
the distribution~$P_\pm$, so that we have sums of operator products of the form
\beq \label{opform}
C_1 \, \B\,C_2 \,\B\, \cdots \,\B\, C_{k+1} \qquad \text{with} \qquad
C_l \in \{s_m, P_+, P_-\} \:.
\eeq

\begin{Def} \label{defhec} An operator product of the form~\eqref{opform} is
a {\bf{low-energy contribution}} if the factors~$C_l$ are all in the set~$\{s_m, P_+\}$ or
are all in~$\{s_m, P_-\}$. Conversely, if the factors~$C_l$ involve both~$P_+$ and~$P_-$,
then the operator product is a {\bf{high-energy contribution}}.
\end{Def}

\begin{Prp} \label{prphec}
Every high-energy contribution is a smooth function on~$M \times M$.
\end{Prp}
\Proof The proposition is obtained by a straightforward adaptation of~\cite[Proof of Theorem~3.4]{light}.
More precisely, following the arguments at the beginning of~\cite[Proof of Theorem~3.4]{light}, it suffices to
consider an operator product of the form
\[ P_+\, \B\, C_{n-1} \,\B \cdots \B\, C_1 \,\B\, P_- \:. \]
Now one can proceed inductively as explained after~\cite[eq.~(3.30)]{light}.
\QED

Hence the singularities of the fermionic projector in position space are determined exclusively by the
low-energy contributions. The singularity structure is described efficiently by the
light-cone expansion (for details see~\cite{firstorder, light} or~\cite[\S2.5]{PFP}),
which is an important tool in the analysis of the continuum limit (see for example~\cite{sector}).
\begin{Def}
A distribution~$A(x,y)$ on~$M \times M$ is
of the order~${\mathcal{O}}((y-x)^{2p})$, $p \in \Z$, if the product
\[ (y-x)^{-2p} \: A(x,y) \]
is a regular distribution (i.e.\ a locally integrable function).
An expansion of the form
\beq
A(x,y) = \sum_{j=g}^{\infty} A^{[j]}(x,y) \label{l:6a}
\eeq
with $g \in \Z$ is called {\bf{light-cone expansion}} if 
if the distributions $A^{[j]}(x,y)$ are of the order
${\mathcal{O}}((y-x)^{2j})$, and if $A$ is approximated by the partial sums
in the sense that for all $p \geq g$,
\beq \label{l:6b}
A(x,y) - \sum_{j=g}^p A^{[j]}(x,y) \qquad {\text{is of the order~${\mathcal{O}}((y-x)^{2p+2})$}}\:.
\eeq
\end{Def} \noindent
The parameter~$g$ gives the leading order of the singularity of~$A(x,y)$ on the light cone.
We point out that we do not demand that the infinite series in~\eqref{l:6a} converges. Thus, similar
to a formal Taylor series, the series in~\eqref{l:6a} is defined via the approximation by the
partial sums~\eqref{l:6b}.

As a simple example for a light-cone expansion, we consider the fermionic projector of
the vacuum~\eqref{Fourier1} and~\eqref{Fourier2}. For clarity, we first leave out the Dirac matrices.
The resulting Fourier integral
\beq \label{Tadef}
T_{m^2}(x,y) := \int \frac{d^4k}{(2 \pi)^4} \: \delta(k^2-m^2)\: \Theta(-k^0) \:e^{-ik(x-y)}
\eeq
can be computed in terms of Bessel functions. Expanding these Bessel functions, one obtains
(see also~\cite[Section~3]{light})
\begin{align*}
T_{m^2}(x,y) &= -\frac{1}{8 \pi^3}
\left( \frac{\text{PP}}{(y-x)^2} \:+\: i \pi \delta \big( (y-x)^2 \big) \: \epsilon \big( (y-x)^0 \big) \right) \\
&\quad +\frac{m^2}{32 \pi^3}\sum_{j=0}^\infty \frac{(-1)^j}{j! \: (j+1)!} \: \frac{\big( m^2 (y-x)^2 \big)^j}{4^j} \\
&\qquad\qquad \times  \Big( \log \big|m^2 (y-x)^2 \big| + c_j
+ i \pi \:\Theta \big( (y-x)^2 \big) \:\epsilon\big( (y-x)^0 \big) \Big)
\end{align*}
with suitable real coefficients~$c_j$. Here ``PP'' denotes the principal value of the integral
(and~$\Theta$ and~$\epsilon$ are again the Heaviside and the step function).
Due to the factor~$(y-x)^{2j}$, this clearly is a light-cone expansion.
The term with the leading singularity becomes integrable after multiplying by~$(y-x)^2$,
showing that~$g=-1$. In view of the factor~$(\slashed{k}+m)$ in~\eqref{Fourier1}, the
light-cone expansion of~$P_m(x,y)$ is obtained by
applying to~$T_{m^2}$ the differential operator~$i \Pdd+m$.
This can be computed term by term in a straightforward manner. Since differentiation
increases the order of the singularity by one, we thus obtain a light-cone expansion
of the form~\eqref{l:6a} with~$g=-2$.

The following theorem makes a general statement on the structure of the light-cone expansion of the
fermionic projector in the presence of an external potential.
\begin{Thm} \label{thmlight} Every contribution to the perturbation expansions
of~$P^\sea$, $P^\sea_\res$, $P^\sea_\ret$ and~$P^\sea_\text{\rm{F}}$ has a light-cone expansion
of the form~\eqref{l:6a}. These {\bf{light-cone expansions}} are {\bf{causal}} in the following sense:
\begin{itemize}
\item[(i)] Every~$A^{[j]}$ is smooth away from the light cone,
\beq \label{Asmooth}
A^{[j]} \in C^\infty \big( \{ (x,y) \in M \times M \:|\: (x-y)^2 \neq 0 \} \big) \:.
\eeq
\item[(ii)] Every~$A^{[j]}$ can be decomposed into a singular and a smooth part,
\[ A^{[j]} = A^{[j]}_\text{\rm{sing}} + A^{[j]}_\text{\rm{reg}} \qquad \text{with} \qquad
A^{[j]}_\text{\rm{reg}} \in C^\infty(M \times M)\:, \]
where the singular part~$A^{[j]}_\text{\rm{sing}}(x,y)$ only depends on~$\B$ 
and its derivatives along the line segment
\[ \overline{xy} = \{ (1+\tau) \,x + \tau y \:|\: 0 \leq \tau \leq 1\} \:. \]
\end{itemize}
\end{Thm} \noindent
In order to explain this notion of ``causality,'' we first point out that if~$x$ and~$y$
are causally separated, then the above line segment is inside the ``causal diamond''
\beq \label{diamond}
\big( J_x^\vee \cap J_y^\wedge \big) \cup \big( J_y^\vee \cap J_x^\wedge \big) \:,
\eeq
where~$J_x^\vee$ and~$J^\wedge_x$ denote the closed future and past light cones centered at~$x$,
\begin{align*}
J_x^\vee &= \{ y \in M \,|\, (y-x)^2 \geq 0,\;
(y^0-x^0) \geq 0 \} \\
J_x^\wedge &= \{ y \in M \,|\, (y-x)^2 \geq 0, \;
(y^0-x^0) \leq 0 \} \:.
\end{align*}
On the other hand, if~$x$ and~$y$ are space-like separated, then~$A^{[j]}(x,y)$
is smooth according to~\eqref{Asmooth}. Thus the above theorem states that the
singularities of the fermionic projector propagate on the light cone and
depend causally on~$\B$.
\Proof[Proof of Theorem~\ref{thmlight}]
The perturbation expansion for the advanced and retarded Green's functions, \eqref{series-scaustilde},
is strictly causal in the sense that it depends on~$\B$ only inside the causal diamond~\eqref{diamond}.
By~\eqref{def-ktil}, the same is true for the causal fundamental solution~$\tilde{k}_m$.
The light-cone expansion of these distributions can be carried out most conveniently
with an iterative construction which involves an expansion in the mass parameter
(see~\cite[Lemma~3.1]{firstorder} and~\cite[Section~2]{light}).
The clue for getting the connection to the light-cone expansions of other operator products
is to perform a suitable expansion in momentum space (as worked out in first order
perturbation theory in~\cite[Section~3]{firstorder}).
Then the so-called residual argument (cf.~\cite[Section~3.1]{light}) shows that the distribution~$\tilde{p}_m^\res$
obtained from~$\tilde{k}_m$ by the replacements~\eqref{r3} and~\eqref{r4}
is also causal in the sense that it has the properties~(i) and~(ii) in the statement of the theorem.
Our task is to show that all the other operators have the same light-cone expansions
as either~$\tilde{k}_m$ or~$\tilde{p}_m^\res$. In view of Proposition~\ref{prphec}, it suffices to show
that all the expansions
\begin{align*}
&\tilde{k}_m^\res - \tilde{k}_m \:, &\hspace*{-1cm} &\tilde{k}^\text{\rm{F}}_m - \tilde{k}_m \\
&\tilde{p}_m - \tilde{p}_m^\res \:, &\hspace*{-1cm} &\tilde{p}^\ret_m - \tilde{p}_m^\res
\end{align*}
only involve high-energy contribution in the sense of Definition~\ref{defhec}.

We next show that the expansions~$\tilde{k}^\res_m - \tilde{k}_m$
and~$\tilde{p}_m - \tilde{p}_m^\res$ only involve high-energy contributions.
To this end, we consider the contour representation of~$\tilde{p}_m^\res$ and~$\tilde{k}_m^\res$
(cf.~\eqref{fres} and~\eqref{ktmres}),
\beq \label{pkres}
\tilde{p}_m^\res = -\frac{1}{2 \pi i} \left( \ointctrclockwise_{\Gamma_+} + \ointctrclockwise_{\Gamma_-} \right)
\tilde{R}_\lambda\: d\lambda \:, \qquad
\tilde{k}_m^\res = -\frac{1}{2 \pi i} \left( \ointctrclockwise_{\Gamma_+} - \ointctrclockwise_{\Gamma_-} \right)
\tilde{R}_\lambda\: d\lambda \:.
\eeq
In order to compute the low-energy contributions, we first substitute the Neumann series~\eqref{tR}.
Then we only take into account the contributions where either for all factors~$R_\lambda$
we consider the poles at~$\lambda=0,1$ (giving the operator products involving~$P_+$)
or for all factors~$R_\lambda$ we consider the poles at~$\lambda=0,-1$ (giving the operator
products involving~$P_-$). In view of~\eqref{pkres}, the low-energy contribution
of~$\tilde{k}_m^\res$ is obtained from that of~$\tilde{p}_m^\res$ by flipping the sign
of those operator products which involve~$P_-$.
Next, since the operator~$\tilde{p}_m^\res$ is symmetric and every factor~$p_m$
or~$k_m$ comes with a factor~$i$, we know that the above operator products all involve
an odd number of factors~$P_-$. Hence the low-energy contribution
of~$\tilde{k}_m^\res$ is also obtained from that of~$\tilde{p}_m^\res$ by flipping the sign
of each factor~$P_-$.
This shows that under the replacements~$k_m \longleftrightarrow p_m$,
the operator~$\tilde{p}^\res_m$ transforms to
\[ \tilde{p}^\res_m \longrightarrow \tilde{k}^\res_m + \text{(high-energy contributions)} \:. \]
The claim now follows because under the same replacements,
we have the transformations~\eqref{r5} and~\eqref{r6}.

It remains to consider the expansions~$\tilde{k}^\text{\rm{F}}_m - \tilde{k}_m$
and~$\tilde{p}^\ret_m - \tilde{p}_m^\res$. These expansions  are obtained from each other by
applying the replacement rules~\eqref{r2}--\eqref{r4} (cf.~\eqref{r5}
and~\eqref{r8}). In view of this symmetry, it suffices to consider the
expansion~$\tilde{k}^\text{\rm{F}}_m - \tilde{k}_m$; the proof for~$\tilde{p}^\ret_m - \tilde{p}_m^\res$
is then immediately obtained by applying the replacements~$p_m \longleftrightarrow k_m$.
Using the definition~\eqref{kfmdef} together with the explicit formulas~\eqref{Updef},
we can substitute~\eqref{spmdec} and multiply out to obtain operator products involving
factors~$s_m$ and~$p_m$ as well as one factor~$k_m$. More precisely,
\beq \label{kFmex}
\tilde{k}^\text{\rm{F}}_m =
\sum_{\alpha, \beta=0}^{\infty} (i\pi)^\alpha \,(-i \pi)^\beta \;b_m^< \,(p_m b_m)^\alpha \,
k_m \,(b_m p_m)^\beta \,b_m^>\:,
\eeq
where the factors~$b_m^\bullet$ are again given by~\eqref{bmdef}.
Computing modulo high-energy contributions, we may replace pairs of factors~$p_m$
by pairs of factors~$k_m$. If~$\alpha+\beta$ is odd, this can be done iteratively until
we end up with operator products involving exactly one factor~$p_m$, i.e.\ which are of the form
\beq \label{term}
(i\pi)^{\alpha'} \,(-i \pi)^{\beta'} \;b_m^< \,(k_m b_m)^{\alpha'} p_m\, (b_m k_m)^{\beta'} \,b_m^>
\qquad \text{with~$\alpha'+\beta'$ odd}\:.
\eeq
Moreover, again computing modulo high-energy contributions, we may exchange two
factors~$p_m$ and~$k_m$, which means in~\eqref{term} that the factor~$p_m$ can be
brought to an arbitrary position. Taking the adjoint of~\eqref{term} and bringing the factor~$p_m$
back to the old position, we obtain minus~\eqref{term}. This shows that all terms
with~$\alpha+\beta$ odd cancel.

In the remaining case when~$\alpha+\beta$ is even, we may
replace all factors~$p_m$ in~\eqref{kFmex} by~$k_m$. We thus obtain
\begin{align*}
\tilde{k}^\text{\rm{F}}_m =
\sum_{\stackrel{\alpha, \beta=0}{\text{$\alpha+\beta$ even}}}^{\infty} (i\pi)^{\alpha+\beta}\, (-1)^\beta
&\;b_m^<\, k_m\, (b_m k_m)^{\alpha+\beta}\, b_m^> \\[-1.7em]
&+ \text{(high-energy contributions)}\:.
\end{align*}
For fixed~$2p:=\alpha+\beta$, we need to sum over the combinations
\[ (\alpha, \beta) = (0,2p), (1, 2p-1), \ldots, (2p,0)\:. \]
Of these~$2p+1$ combinations, $p+1$ contribute with a plus sign, whereas~$p$
of them give a minus sign. Adding up, we obtain precisely the formula~\eqref{kex} for~$\tilde{k}_m$.
\QED

It is quite remarkable that all our perturbation expansions have causal light-cone expansions.
It is not known whether there is a simple criterion to decide which operator products
have a causal light-cone expansion. Instead, we finally give a simple example for a
perturbation expansion that has a light-cone expansion which is {\em{not causal}} in the above
sense. To this end, we consider the perturbation series
\beq \label{Pfalse}
\tilde{P}^\sea = \sum_{\alpha, \beta=0}^\infty
(-s_m^- \B)^\alpha P_- (-\B s_m^-)^\beta \:.
\eeq
This perturbation series differs from our expansion with the Feynman propagator~\eqref{PFdef}
in that we are using the same propagator~$s_m^-$ on the left and on the right.
As a consequence, the operator~$\tilde{P}^\sea_\text{\rm{F}}$ is not symmetric.
But if we are willing to give up symmetry, then the above series is another
possible perturbation expansion for the fermionic projector. To first order in~$\B$, we get the
contribution
\[ -s_m^- \B P_- - P_- \B s_m^- \:. \]
Using~\eqref{spmdec}, this can be rewritten as
\[ \big( -s_m \B P_- - P_- \B s_m \big) + i \pi \big(  p_m \B P_- + P_- \B p_m^- \big) \:. \]
The terms in the first bracket coincides precisely with the first order perturbation
of the fermionic projector with mass or spatial normalization.
The terms in the second bracket, however, are a consequence of the specific form of
the perturbation expansion~\eqref{Pfalse}.
As worked out in detail in~\cite[Lemmas~F.3 and~F.4]{PFP},
its light-cone expansion involves unbounded line integrals
which violate the property~(ii) in Theorem~\ref{thmlight}.

\section{Fermion Loops} \label{secloop}
As explained in~\cite{qft}, fermionic loop diagrams are obtained in the fermionic projector approach
by considering the fermionic projector~$P(x,y)$ for~$x=y$ after subtracting suitable singular contributions
(see~\cite[eq.~(2.13)]{qft})
\beq \label{Psing}
P(x,x) - {\text{(singular contributions)}} \:.
\eeq
Here we do not enter the analysis of the singular contributions
(for details see~\cite[Section~6 and~7]{sector}). Instead, we merely consider the
contributions to~$P(x,x)$ with a simple ultraviolet regularization.
Our goal is to show that certain contributions to the perturbation expansion
vanish by symmetry, in generalization of Furry's theorem in standard quantum field theory.
We anticipate that this symmetry argument also applies to the singular contributions in~\eqref{Psing}
(independent of the detailed regularization method), implying that the corresponding
diagrams do not contribute to the fermion loops~\eqref{Psing}.

\subsection{A Generalized Furry Theorem}
Recall the definitions of the following Green's functions (see~\eqref{spmdef} and~\eqref{sadret}, \eqref{sdef}):
\begin{align}
s_m^\pm(k) &= \lim_{\varepsilon \searrow 0} \frac{\slashed{k}+m}{k^2 - m^2 \mp i \varepsilon} \label{spm} \\
s^\vee_m(k) &= \lim_{\varepsilon \searrow 0}
\frac{\slashed{k} + m}{k^{2}-m^{2}-i \varepsilon k^{0}} \\
s^\wedge_m(k) &= \lim_{\varepsilon \searrow 0}
\frac{\slashed{k} + m}{k^{2}-m^{2}+i \varepsilon k^{0}} \\
s_m(k) &= \frac{1}{2}\, \big(s_m^\vee+s_m^\wedge \big)(k)
= \frac{1}{2}\, \big(s_m^- + s_m^+ \big)(k)
= (\slashed{k} + m)\: \frac{\text{PP}}{k^{2}-m^{2}}  \label{s}
\intertext{(where ``PP'' again denotes the principal value).
Taking differences of these Green's functions, we obtain the
fundamental solutions (cf.~\eqref{kdef}, \eqref{ksrel} and~\eqref{pdef}, \eqref{spmdec}),}
p_m(k) &= \frac{1}{2 \pi i} \,\big(s_m^- - s_m^+ \big)(k) = (\slashed{k}+m)\:\delta(k^2-m^2) \label{p} \\
k_m(k) &= \frac{1}{2 \pi i} \,\big(s_m^\vee-s_m^\wedge \big)(k) = (\slashed{k}+m)\:\delta(k^2-m^2)\:
\epsilon(k^0)\:. \label{k}
\end{align}
Thus we may decompose any Green's function in terms of~$s_m$ and a fundamental solution,
\begin{align*}
s_m^\pm &= s_m \pm i \pi\, p_m \\
s_m^\vee &= s_m + i \pi\, k_m \\
s_m^\wedge &= s_m - i \pi\, k_m \:.
\end{align*}

We consider products involving of the form
\[ C_0(x_0, x_1) \,\B_1\, C_1(x_1, x_2) \, \cdots \, C_{n-1}(x_{n-1}, x_n)\,
\B_n\, C_n(x_n, x_0) \:, \]
where the factors~$C_0, \ldots, C_n$ represent any of the above distributions,
and the factors~$\B_1, \ldots \B_n$ stand for an odd combination of Dirac matrices, i.e.
\[ \B_k = \slashed{A}_k + \gamma^5 \slashed{B}_k \:. \]
In order for these products to be well-defined, one should consider the regularized
distributions, in the simplest case by taking the above formulas for fixed~$\varepsilon>0$.
All our arguments apply just as well to the regularized product. For ease in notation, we
prefer to work with the unregularized distributions.
\begin{Prp} \label{prpadv}
If the factors~$C_0, \ldots, C_n$ are all the advanced Green's functions,
then the operator product vanishes:
\[ s^\vee(x_0, x_1) \,\B_1\, s^\vee(x_1, x_2) \, \cdots \, s^\vee(x_{n-1}, x_n)\, \B_n\, s^\vee(x_n, x_0) = 0 \:. \]
The same holds if all the factors are retarded Green's functions. 
\end{Prp}
\Proof The advanced Green's functions are non-zero only if~$x_1$ lies in the future of~$x_0$,
$x_2$ lies in the future of~$x_1$, \ldots, and~$x_0$ lies in the future of~$x_n$. This is impossible.
We note that the causality of the Green's function also holds with regularization,
as one sees by taking the Fourier transform of~$(k^2-m^2+i \varepsilon k^0)^{-1}$ with residues.

The argument for the retarded Green's functions is similar.
\QED
The following result generalizes Furry's theorem in standard quantum field theory
(see for example~\cite[p.~331ff]{bogo-shir2}) to more general distributions than the Feynman propagator.
\begin{Thm} \label{thmfurry} Suppose that the factors~$C_0, \ldots, C_n$ are a selection of
any of the distributions~$s_m^\pm$, $s_m$, $p_m$ or~$k_m$. Then
\begin{align*}
\Tr &\big( \B_0\, C_0(x_0, x_1) \,\B_1\, C_1(x_1, x_2) \, \cdots \,
C_{n-1}(x_{n-1}, x_n)\, \B_n\, C_n(x_n, x_0) \big) \\
&= (-1)^{n+k+1}
\Tr \big( C_n(x_0, x_n) \,\B_n \,C_{n-1}(x_n, x_{n-1})\, \cdots\,C_1(x_2, x_1) \,\B_1\,C_0(x_1, x_0)\, \B_0 \big)\:,
\end{align*}
where~$k$ denotes the number of factors~$k_m$.
\end{Thm}
\Proof Exactly as explained in~\cite[p.~331]{bogo-shir2}, in traces of products of Dirac matrices
we may transpose all matrices according to the transformation
\beq \label{arr1}
\gamma \rightarrow -\gamma^\text{T} \:,\qquad \gamma_{\alpha \beta} \rightarrow - \gamma_{\beta \alpha} \:.
\eeq
From a more abstract point of view, this transformation can also be understood as follows.
Obviously, the matrices~$\gamma^\text{T}$ again satisfy the anti-commutation relations.
Hence they form a representation of the Clifford algebra. Since in dimension four all
irreducible representations are equivalent, there is an invertible matrix~$S$ such that
\[ \gamma = -S \gamma^\text{T} S^{-1}\:. \]
Using this identity for all Dirac matrices, all factors~$S$ and~$S^{-1}$ cancel each other
in the trace of operator products.

Applying the transformation~\eqref{arr1}, 
each factor~$\B_k$ gives a minus sign. In the factors~$C_k$ we transform the Fourier
integral to obtain
\begin{align*}
C(x,y) &= \int  \frac{d^4 k}{(2 \pi)^4}\: (\slashed{k}+m)\: f(k)\: e^{-ik(x-y)} \\
&\rightarrow  \int  \frac{d^4 k}{(2 \pi)^4}\: (-\slashed{k}+m)^\text{T}\: f(k)\: e^{-ik(x-y)} \\
&= \int  \frac{d^4 k}{(2 \pi)^4}\: (\slashed{k}+m)^\text{T}\: f(-k)\: e^{ik(x-y)} 
= \int  \frac{d^4 k}{(2 \pi)^4}\: (\slashed{k}+m)^\text{T}\: f(-k)\: e^{-ik(y-x)} \:.
\end{align*}
Using the explicit formulas~\eqref{spm}, \eqref{s}, \eqref{p} and~\eqref{k}, one
sees that the corresponding function~$f$ is even for~$s_m^\pm$, $s_m$ and~$p_m$,
but is odd for~$k_m$. This gives the transformations
\beq \label{arr2}
\begin{split}
s_m^\pm(x,y) &\rightarrow s_m^\pm(y,x)^\text{T} \:, \qquad \qquad s_m(x,y) \rightarrow s_m(y,x)^\text{T} \:,\\
p_m(x,y) &\rightarrow p_m(y,x)^\text{T} \:, \hspace*{1.5173cm} k_m(x,y) \rightarrow -k_m(y,x)^\text{T} \:.
\end{split}
\eeq
It follows that
\begin{align*}
\Tr& \big( \B_0\, C_0(x_0, x_1) \,\B_1\, C_1(x_1, x_2) \, \cdots \,
C_{n-1}(x_{n-1}, x_n)\, \B_n\, C_n(x_n, x_0) \big) \\
&= \sum_{\stackrel{\alpha_0, \ldots, \alpha_n}{\beta_0, \ldots, \beta_n}}
(\B_0)_{\alpha_0 \beta_0}\, C_0(x_0, x_1)_{\beta_0, \alpha_1} \,(\B_1)_{\alpha_1 \beta_1}
\, C_1(x_1, x_2)_{\beta_1 \alpha_2} \\[-1.5em]
& \hspace*{4cm} \cdots \, C_{n-1}(x_{n-1}, x_n)_{\beta_{n-1} \alpha_n} \, (\B_n)_{\alpha_n \beta_n}\, C_n(x_n, x_0)_{\beta_n \alpha_0} \\
&\overset{(\ast)}{=} (-1)^{n+k+1} \sum_{\stackrel{\alpha_0, \ldots, \alpha_n}{\beta_0, \ldots, \beta_n}}
(\B_0)_{\beta_0 \alpha_0}\, C_0(x_1, x_0)_{\alpha_1, \beta_0} \,(\B_1)_{\beta_1 \alpha_1}
\, C_1(x_2, x_1)_{\alpha_2 \beta_1} \\[-1.5em]
& \hspace*{4cm} \cdots \, C_{n-1}(x_n, x_{n-1})_{\alpha_n \beta_{n-1}} \, (\B_n)_{\beta_n \alpha_n}\, C_n(x_0, x_n)_{\alpha_0 \beta_n} \\
&= (-1)^{n+k+1} \Tr \big( C_n(x_0, x_n)\,\B_n\,C_{n-1}(x_n, x_{n-1})\, \cdots \,
C_1(x_2, x_1) \,\B_1\, C_0(x_1, x_0) \,\B_0 \big) ,
\end{align*}
where in~$(\ast)$ we substituted~\eqref{arr1} and~\eqref{arr2}. This concludes the proof.
\QED
We now apply this theorem to perturbation expansions of the fermionic projector.
\begin{Corollary} \label{corfurry}
Consider the fermionic projector in the presence of an external potential~$\B$
which is odd, i.e.
\beq \label{Bodd}
\B = \slashed{A} + \gamma^5 \slashed{B} \:.
\eeq
Let~$\Delta P^{(n,k)}$ be the contribution to the perturbation expansion of the projector of a fixed order~$n$
which involves~$k$ factors~$k_m$. Then for any
any odd matrix~${\mathscr{U}} = \slashed{u} + \gamma^5 \slashed{v}$,
\[ \Tr \big( {\mathscr{U}}\, \Delta P^{(n,k)}(x,x) \big) = 0 \qquad
\text{if~$n+k$ is even}\:. \]
\end{Corollary}
\Proof Since the perturbation expansion is symmetric under
transpositions of the factors, it follows that~$\Delta P^{(n,k)}$ can be written as a sum of operator products
of the form
\[ C_0\,\B\, C_1\, \cdots \, C_{n-1}\, \B\, C_n 
+ C_n\,\B\, C_{n-1}\, \cdots \, C_1\, \B\, C_0  \:. \]
We now apply Theorem~\ref{thmfurry} to obtain the result.
\QED

\subsection{First Order Loop Diagrams}
We now compute the one-loop contribution in an external potential~$\B$
and simplify the formulas with the help of Furry's theorem.

\begin{Prp} \label{prponeloop}
Consider the fermionic projector in the presence of an external potential~$\B$
which is odd~\eqref{Bodd}.
Then to first order in the external potential, the vectorial and axial one-loop contributions are the same for all
considered perturbation expansions. More precisely, for
any odd matrix~${\mathscr{U}} = \slashed{u} + \gamma^5 \slashed{v}$,
\begin{align*}
\Tr &({\mathscr{U}} \,\Delta P^\sea_\res(x,x)) = \Tr ({\mathscr{U}} \,\Delta P^\sea(x,x)) 
= \Tr ({\mathscr{U}} \,\Delta P^\sea_\ret(x,x)) = \Tr ({\mathscr{U}} \,\Delta P^\sea_\text{\rm{F}}(x,x)) \\
&= -\frac{1}{2} \Tr \Big( \slashed{u} \,\big( p_m \,\B\, s_m + s_m  \,\B\, p_m \big)(x,x) \Big) .
\end{align*}
\end{Prp}
\Proof To first order in the external potential, we have
\begin{align*}
\Delta P^\sea_\res = \Delta P^\text{sea} &= - s_m \,\B\, P_- - P_- \,\B\, s_m \\
\Delta P^\sea_\ret &= -s_m^\wedge \,\B\, P_- - P_- \,\B\, s_m^\vee
= -(s_m - i \pi \,k_m) \,\B\, P_- - P_- \,\B\, (s_m + i \pi \,k_m) \\
&= \Delta P^\text{sea} + i \pi\, \big( k_m \,\B\, P_- - P_- \,\B\, k_m \big) \\
\Delta P^\sea_\text{\rm{F}} &= -s_m^- \,\B\, P_- - P_- \,\B\, s_m^+
= -(s_m - i \pi \,p_m) \,\B\, P_- - P_- \,\B\, (s_m + i \pi \,p_m) \\
&= \Delta P^\text{sea} + i \pi\, \big( p_m \,\B\, P_- - P_- \,\B\, p_m \big)
\end{align*}
(with~$P_-$ according to~\eqref{Ppmdef}).
As a consequence, applying Corollary~\ref{corfurry},
\begin{align*}
\Tr ({\mathscr{U}} \,\Delta P^\sea_\res(x,x)) &= \Tr ({\mathscr{U}} \,\Delta P^\sea(x,x)) = 
-\frac{1}{2} \Tr \Big( {\mathscr{U}} \,\big( p_m \,\B\, s_m + s_m  \,\B\, p_m \big)(x,x) \Big) \\
\Tr ({\mathscr{U}} \,\Delta P^\sea_\ret(x,x)) &=\Tr ({\mathscr{U}} \,\Delta P^\sea(x,x))
+i \pi \Tr \Big( {\mathscr{U}} \,\big( k_m \,\B\, P_- - P_- \,\B\, k_m \big)(x,x) \Big) \\
&= \Tr ({\mathscr{U}} \,\Delta P^\sea(x,x))
-\frac{i \pi}{2} \Tr \Big( {\mathscr{U}} \,\big( k_m \,\B\, k_m - k_m \,\B\, k_m \big)(x,x) \Big) \\
\Tr ({\mathscr{U}} \,\Delta P^\sea_\text{\rm{F}}(x,x)) &=\Tr ({\mathscr{U}} \,\Delta P^\sea(x,x))
+i \pi \Tr \Big( {\mathscr{U}} \,\big( p_m \,\B\, P_- - P_- \,\B\, p_m \big)(x,x) \Big) \\
&= \Tr ({\mathscr{U}} \,\Delta P^\sea(x,x))
+\frac{i \pi}{2} \Tr \Big( {\mathscr{U}} \,\big( p_m \,\B\, p_m - p_m \,\B\, p_m \big)(x,x) \Big) \:.
\end{align*}
This gives the result.
\QED

It is worth noting that this contribution differs from the first-order loop diagram in
standard quantum field theory obtained using the Feynman propagator.
\begin{Lemma} Under the assumptions of Proposition~\ref{prponeloop},
\begin{align}
\Tr ({\mathscr{U}} (s_m^- \,\B\, s_m^-)(x,x)) &= -2 \pi i \Tr \Big( {\mathscr{U}} \, \Delta P^\sea(x,x) \Big) \label{onefeyn1} \\
&\qquad - \pi^2 \Tr \Big( {\mathscr{U}} \,\big( p_m \,\B\, p_m - k_m \,\B\, k_m \big)(x,x) \Big) . \label{onefeyn2}
\end{align}
\end{Lemma}
\Proof
\begin{align*}
s_m^- \,\B\, s_m^- &= (s_m - i \pi\, p_m) \,\B\, (s_m - i \pi\, p_m) \\
&= s_m \,\B\, s_m - i \pi \,\big( p_m \,\B\, s_m + s_m  \,\B\, p_m \big)
- \pi^2 \, p_m\,\B\, p_m \\
s_m^\wedge \,\B\, s_m^\wedge &= (s_m - i \pi\, k_m) \,\B\, (s_m - i \pi\, k_m) \\
&= s_m \,\B\, s_m - i \pi \big( k_m \,\B\, s_m + s_m  \,\B\, k_m \big)
- \pi^2 \, k_m\,\B\, k_m \\
\intertext{Subtracting these identities, we obtain}
s_m^- \,\B\, s_m^- - s_m^\wedge \,\B\, s_m^\wedge &=
- 2 \pi i \big( P_- \,\B\, s_m + s_m  \,\B\, P_- \big)
- \pi^2\, \big( p_m \,\B\, p_m - k_m \,\B\, k_m \big) \:.
\end{align*}
Evaluating the operator products on the diagonal~$(x,x)$, 
the operator product involving~$s_m^\vee$ vanishes according to Proposition~\ref{prpadv}.
This gives the result.
\QED
The term on the right of~\eqref{onefeyn1} was discussed by Dirac~\cite{dirac3},
Heisenberg~\cite{heisenberg2} and computed by Uehling~\cite{uehling} in the static situation.
The term on the left of~\eqref{onefeyn1} was first computed in~\cite{feynmanloop}.
We point out that the term~\eqref{onefeyn2} is a high-energy contribution, which clearly vanishes in the static situation.

\subsection{Second Order Loop Diagrams}
To second order, the contributions to the loop diagram depend on the considered perturbation
expansion. Only for~$P^\sea$ we get zero (in agreement with the usual loop computation
using the Feynman propagator). In all the other perturbation expansions we get
non-trivial high-energy contributions. More precisely, we have the following result.

\begin{Prp} \label{prpsecondloop}
Consider the fermionic projector in the presence of an external potential~$\B$
which is odd~\eqref{Bodd}.
Then for the contribution of second order in the external potential and
any odd matrix~${\mathscr{U}} = \slashed{u} + \gamma^5 \slashed{v}$, one obtains
\begin{align*}
\Tr ({\mathscr{U}} \,\Delta P^\sea(x,x)) =\;& 0 \\
\Tr ({\mathscr{U}} \,\Delta P^\sea_\res(x,x)) =\;& -\frac{\pi^2}{4}
\Tr \Big({\mathscr{U}} \, \big( k_m \,\mathscr{B}\, k_m \,\mathscr{B}\, k_m
-p_m \,\mathscr{B} \,p_m\, \mathscr{B} \,k_m \\
&\hspace*{1.8cm}
+p_m \,\mathscr{B}\, k_m \,\mathscr{B}\, p_m
-k_m \,\mathscr{B}\, p_m \,\mathscr{B}\, p_m \big)(x,x) \Big) \\
\Tr ({\mathscr{U}} \,\Delta P^\sea_\ret(x,x)) =\;&  -\frac{i \pi}{2}
\Tr \Big({\mathscr{U}} \, \big( k_m \,\mathscr{B}\, p_m \,\mathscr{B}\, s_m
+ k_m \,\mathscr{B}\, s_m \,\mathscr{B}\, p_m
- p_m \,\mathscr{B}\, k_m \,\mathscr{B}\, s_m \\
&\quad
- p_m \,\mathscr{B}\, s_m \,\mathscr{B}\, k_m
+ s_m \,\mathscr{B}\, k_m \,\mathscr{B}\, p_m
- s_m \,\mathscr{B}\, p_m \,\mathscr{B}\, k_m \big)(x,x) \Big) \\
\Tr ({\mathscr{U}} \,\Delta P^\sea_\text{\rm{F}}(x,x)) =\;& -\frac{i \pi}{2}
\Tr \Big({\mathscr{U}} \, \big( k_m \,\mathscr{B}\, p_m \,\mathscr{B}\, s_m
+ k_m \,\mathscr{B}\, s_m \,\mathscr{B}\, p_m
- p_m \,\mathscr{B}\, k_m \,\mathscr{B}\, s_m \\
&\quad
- p_m \,\mathscr{B}\, s_m \,\mathscr{B}\, k_m
+ s_m \,\mathscr{B}\, k_m \,\mathscr{B}\, p_m
- s_m \,\mathscr{B}\, p_m \,\mathscr{B}\, k_m \big)(x,x) \Big) \\
&-\frac{\pi^2}{2} \Tr \Big({\mathscr{U}} \, \big( 
k_m \,\mathscr{B}\, k_m \,\mathscr{B}\, k_m
-k_m \,\mathscr{B}\, p_m \,\mathscr{B}\, p_m \\
&\hspace*{1.8cm}
-p_m \,\mathscr{B}\, k_m \,\mathscr{B}\, p_m
+p_m \,\mathscr{B}\, p_m \,\mathscr{B}\, k_m \big)(x,x) \Big) .
\end{align*}
\end{Prp}
\Proof We begin with~$P^\sea$. The contribution of second order in the external potential is given
by (cf.\ the formula for~$\tilde{t}$ in~\cite[Appendix~A]{grotz}),
\begin{align*}
\Delta P^\sea &= \Big( P_- \,\mathscr{B} \,s_m \,\mathscr{B}\, s_m + s_m \,\mathscr{B} \,P_- \,\mathscr{B}\, s_m
+ s_m \,\mathscr{B} \,s_m \,\mathscr{B}\, P_- \Big) + \frac{\pi^2}{2}\: k_m \,\mathscr{B} \, k_m \,\mathscr{B}\, k_m \\
& \qquad +\frac{\pi^2}{4}\Big(k_m \,\mathscr{B}\, p_m \,\mathscr{B}\, k_m
-k_m \,\mathscr{B} \,k_m\, \mathscr{B} \,p_m
-p_m \,\mathscr{B}\, k_m \,\mathscr{B}\, k_m
-p_m \,\mathscr{B}\, p_m \,\mathscr{B}\, p_m\Big) \:.
\end{align*}
Applying Corollary~\ref{corfurry}, in the loops all contributions involving an even number of factors~$k_m$ vanish.
Thus
\begin{align}
\Tr ({\mathscr{U}} \,\Delta P^\sea(x,x)) &= 
-\frac{1}{2} \Tr \Big({\mathscr{U}} \, \big(
k_m \,\mathscr{B} \,s_m \,\mathscr{B}\, s_m + s_m \,\mathscr{B} \,k_m \,\mathscr{B}\, s_m
+ s_m \,\mathscr{B} \,s_m \,\mathscr{B}\, k_m \big)(x,x) \Big) \nonumber \\
&\qquad + \frac{\pi^2}{2} \Tr \Big({\mathscr{U}} \, \big(k_m \,\mathscr{B} \, k_m \,\mathscr{B}\, k_m
\big)(x,x) \Big) \:. \label{DelP}
\end{align}
On the other hand, Proposition~\ref{prpadv} yields
\begin{align*}
0 &= \big( s_m^\wedge \,\B\, s_m^\wedge \,\B\, s_m^\wedge -
s_m^\vee \,\B\, s_m^\vee \,\B\, s_m^\vee \big)(x,x) \\
&= -2 i \pi \big( k_m \,\mathscr{B} \,s_m \,\mathscr{B}\, s_m + s_m \,\mathscr{B} \,k_m \,\mathscr{B}\, s_m
+ s_m \,\mathscr{B} \,s_m \,\mathscr{B}\, k_m \big)(x,x) \\
&\qquad + 2 i \pi^3 \big(k_m \,\mathscr{B} \, k_m \,\mathscr{B}\, k_m \big)(x,x) \:.
\end{align*}
Using these relations, all the terms in~\eqref{DelP} cancel.

The result for~$\Delta P^\sea_\res$ follows similarly from the formula (see~\cite[Appendix~A]{grotz}),
\begin{align*}
\Delta P^\sea_\res &= \Big( P_- \,\mathscr{B} \,s_m \,\mathscr{B}\, s_m
+ s_m \,\mathscr{B} \,P_- \,\mathscr{B}\, s_m + s_m \,\mathscr{B} \,s_m \,\mathscr{B}\, P_- \Big)
- \frac{\pi^2}{2}\: p_m \,\mathscr{B} \, p_m \,\mathscr{B}\, p_m \\
& \qquad +\frac{\pi^2}{4}\Big(k_m \,\mathscr{B}\, k_m \,\mathscr{B}\, k_m
+p_m \,\mathscr{B} \,p_m\, \mathscr{B} \,k_m
-p_m \,\mathscr{B}\, k_m \,\mathscr{B}\, p_m
+k_m \,\mathscr{B}\, p_m \,\mathscr{B}\, p_m\Big) \:.
\end{align*}
The computation for~$\Delta P^\sea_\ret$ and~$\Delta P^\sea_\text{\rm{F}}$ is analogous.
\QED

\appendix

\section{The Leading Orders of the Perturbation Expansions} \label{appA}
We now give explicit formulas up to third order of the perturbation series
with spatial and mass normalization.
In these computations, it is convenient to use in addition to~\eqref{rules} the
computation rules
\[ s \sdot p = s \sdot k = p \sdot s = k \sdot s = 0 \qquad \text{and} \qquad
s \sdot s = \pi^2 \, p \:, \]
which are obtained from~\cite[Lemma~]{grotz} by omitting all contributions involving
principal values (omitting the principal values is admissible because all operator products
considered before~\eqref{sdotdef} give rise to a factor~$\delta(m-m')$,
which implies that all principal values cancel each other in all our computations).
All explicit computations were carried out with the help of
the {\textsf{Mathematica}} package {\textsf{BasicCausal.m}}\footnote{This package is available as an
ancillary file on the arXiv.}.

We first give the expansions of the fermionic projector with spatial and mass normalization.
These formulas were first given in~\cite[Appendix~A]{grotz} (albeit without analyzing the spatial
normalization), and we here restate them for the sake of completeness.
The operators~$\tilde{k}$ and~$\tilde{p}$ in~\eqref{ktildef} and~\eqref{ptm} have the expansions
\begin{align*}	
\tilde{k} =\;&k-s\mathscr{B}k-k\mathscr{B}s+k\mathscr{B}s\mathscr{B}s+s\mathscr{B}k\mathscr{B}s+s\mathscr{B}s\mathscr{B}k-\pi^2k\mathscr{B}k\mathscr{B}k\\
&-k\mathscr{B}s\mathscr{B}s\mathscr{B}s-s\mathscr{B}k\mathscr{B}s\mathscr{B}s-s\mathscr{B}s\mathscr{B}k\mathscr{B}s-s\mathscr{B}s\mathscr{B}s\mathscr{B}k\\
&+\pi^2\Big(s\mathscr{B}k\mathscr{B}k\mathscr{B}k+k\mathscr{B}s\mathscr{B}k\mathscr{B}k+k\mathscr{B}k\mathscr{B}s\mathscr{B}k+k\mathscr{B}k\mathscr{B}k\mathscr{B}s\Big)+\O(\mathscr{B}^4) \\
\tilde{p} =\;& p-s\mathscr{B}p-p\mathscr{B}s+p\mathscr{B}s\mathscr{B}s+s\mathscr{B}p\mathscr{B}s+s\mathscr{B}s\mathscr{B}p\\
&+\frac{\pi^2}{2}\Big(-p\mathscr{B}k\mathscr{B}k+k\mathscr{B}p\mathscr{B}k-k\mathscr{B}k\mathscr{B}p-p\mathscr{B}p\mathscr{B}p\Big)\\
&-p\mathscr{B}s\mathscr{B}s\mathscr{B}s-s\mathscr{B}p\mathscr{B}s\mathscr{B}s-s\mathscr{B}s\mathscr{B}p\mathscr{B}s-s\mathscr{B}s\mathscr{B}s\mathscr{B}p\\
&+\frac{\pi^2}{2}\Big(s\mathscr{B}p\mathscr{B}p\mathscr{B}p+p\mathscr{B}s\mathscr{B}p\mathscr{B}p+p\mathscr{B}p\mathscr{B}s\mathscr{B}p+p\mathscr{B}p\mathscr{B}p\mathscr{B}s\\
&\qquad\quad+p\mathscr{B}s\mathscr{B}k\mathscr{B}k-s\mathscr{B}k\mathscr{B}p\mathscr{B}k+s\mathscr{B}k\mathscr{B}k\mathscr{B}p+k\mathscr{B}s\mathscr{B}k\mathscr{B}p-k\mathscr{B}p\mathscr{B}k\mathscr{B}s\\
&\qquad\quad+p\mathscr{B}k\mathscr{B}s\mathscr{B}k+p\mathscr{B}k\mathscr{B}k\mathscr{B}s-k\mathscr{B}p\mathscr{B}s\mathscr{B}k+k\mathscr{B}k\mathscr{B}p\mathscr{B}s+s\mathscr{B}p\mathscr{B}k\mathscr{B}k\\
&\qquad\quad-k\mathscr{B}s\mathscr{B}p\mathscr{B}k+k\mathscr{B}k\mathscr{B}s\mathscr{B}p\Big)+\O(\mathscr{B}^4) \:.
\end{align*}
The operators $\tilde{p}^{\,\res}$ and~$\tilde{k}^{\,\res}$ in \eqref{ptmres} and~\eqref{ktmres}
are obtained by the replacements~\eqref{r2}--\eqref{r4}, \eqref{r5} and~\eqref{r6}.
The fermionic projectors with spatial normalization and mass normalization are
given by~\eqref{Ppk}.

We come to the computation of the unitary perturbation flow. Before stating our results,
we point out to a complication when computing products in~\eqref{Usdef}:
The factors~$\tilde{p}^\res$ in~\eqref{Usdef} involve different external potentials.
As a consequence, instead of~\eqref{bbprod} we need the more general computation rule
\begin{align}
k_m \,b_m^>[\B] \: b_{m'}^<[\tilde{B}]\: k_{m'} &= \delta(m-m') \Big( p_m
+ \pi^2\,k_m\: b_m[\B]\: p_m\: b_m[\tilde{B}]\: k_m \Big) \\
&\qquad + \frac{\text{PP}}{m-m'} \: k_m \,\big( 1 - b_m^>[\B] \big)\: (\B - \tilde{\B})\: 
\big(1 - b_{m'}^<[\tilde{B}])\: k_{m'} \:, \label{BmtB}
\end{align}
where the square brackets clarify the dependence on the external potential and ``PP''
again denotes the principal value (this rule is obtained by a straightforward computation using the
multiplication rules in~\cite[Lemma~2.1]{grotz}).
The summand~\eqref{BmtB} exhibits the fact that the solution space
for the potential~$\B$ and mass~$m$ is in general not orthogonal to the solution space
for another potential~$\B'$ and a different mass~$m' \neq m$.
In the usual description of adiabatic processes in Hilbert spaces, the corresponding contributions
to a product of the form~\eqref{Usdef} decay like~$1/N$ and thus vanish in the
limit~$N \rightarrow \infty$ (this is precisely the reason why~\eqref{Usdef} is unitary).
In our perturbative description, the terms~$(\B-\tilde{\B})$ in~\eqref{BmtB} also gives the desired factor~$1/N$.
However, since we are here working in an indefinite inner product space,
proving that the summand~\eqref{BmtB} vanishes in~\eqref{Usdef} in the limit~$N \rightarrow \infty$
requires methods which we do not want to enter here
(more precisely, one would have to use ``completeness relations'' obtained by integrating
the mass parameter on a contour~$C_\varepsilon$ as introduced in~\cite[Section~5]{grotz}).
Instead, we simply make the summand~\eqref{BmtB}
vanish by working in addition to~\eqref{rules} with the multiplication rule
\beq \label{rules2}
s \sdot s = \pi^2 \: p
\eeq
(which corresponds to and harmonizes with~\eqref{rule5} if a spatial normalization is used).
Then the unitary perturbation flow with mass normalization (cf.~\eqref{Usdef})
can be computed in a straightforward manner. One obtains the expansion
\begin{align*}
U_\res =\;& p - s \mathscr{B} p + s \mathscr{B} s \mathscr{B} p 
-\frac{\pi^2}{4} \Big( k \mathscr{B} k \mathscr{B} p - p \mathscr{B} k \mathscr{B} k
+ 2\, p \mathscr{B} p \mathscr{B} p \Big) \\
&+\frac{\pi^2}{12} \Big(
k \mathscr{B} p \mathscr{B} s \mathscr{B} k -k \mathscr{B} s \mathscr{B} p \mathscr{B} k
+ 7\, p \mathscr{B} p \mathscr{B} s \mathscr{B} p
+ 5\, p \mathscr{B} s \mathscr{B} p \mathscr{B} p
+ 6\, s \mathscr{B} p \mathscr{B} p \mathscr{B} p \Big) \\
&+\frac{\pi^2}{4} \Big(
k \mathscr{B} k \mathscr{B} s \mathscr{B} p + k \mathscr{B} s \mathscr{B} k \mathscr{B} p
+ s \mathscr{B} k \mathscr{B} k \mathscr{B} p \Big) \\
&-\frac{\pi^2}{4} \Big(
p \mathscr{B} k \mathscr{B} s \mathscr{B} k + p \mathscr{B} s \mathscr{B} k \mathscr{B} k
+ s \mathscr{B} p \mathscr{B} k \mathscr{B} k \Big) +\O(\mathscr{B}^4) \:.
\end{align*}
As is verified by a straightforward computation, the formulas of Proposition~\ref{prpflow}
all hold, giving an a-posteriori justification of the rule~\eqref{rules2}.
The perturbation flow with spatial normalization~$U$ (see Proposition~\ref{prpflowspatial})
is obtained by applying the replacement rules~\eqref{r2}--\eqref{r4}.

We finally derive~\eqref{Ures2}. For a closed loop~$(\B(\tau))_{0 \leq \tau \leq 1}$ with~$\B(0)=\B(1)=0$,
the Dyson series~\eqref{Uresex} simplifies with the help of the rules~\eqref{rules}
and the fact that~$\tilde{p}^\res(0)=p=\tilde{p}^\res(1)$ to
\[ U_\res^*(1) = p + p \sdot \int_0^1 ds_1 \int_0^{s_1} ds_2\:
(\tilde{p}^\res)'(s_2) \sdot \,(\tilde{p}^\res)'(s_1)  + \O(\B^3) \:. \]
Now we insert the first order expansion
\[ (\tilde{p}^\res)'(\tau) =\; - s\, \mathscr{B}'(\tau)\, p-p \,\mathscr{B}'(\tau)\, s + \O(\B^2) \:. \]
Again using the rules~\eqref{rules} as well as~\eqref{rules2}, we obtain
\begin{align*}
U_\res^*(1) &= p + p \sdot \int_0^1 ds_1 \int_0^{s_1} ds_2\;
\big( p \,\B'(s_2)\, s \big) \sdot \big( s \,\B'(s_1)\, p \big) + \O(\B^3) \\
&= p + \pi^2 \int_0^1 ds_1 \int_0^{s_1} ds_2\;
p \,\B'(s_2)\, p \,\B'(s_1)\, p + \O(\B^3) \:.
\end{align*}
Carrying out the integral over~$s_2$ and using that~$\B(0)=0$, we obtain~\eqref{Ures2}.

\Thanks{{{\em{Acknowledgments:}}
We would like to thank the referee for helpful comments on the manuscript.
F.F.\ is grateful to the Max Planck Institute for Mathematics in the Sciences, Leipzig, for its hospitality.}}


\begin{thebibliography}{10}

\bibitem{berry}
M.V. Berry, \emph{Quantal phase factors accompanying adiabatic changes}, Proc.
  Roy. Soc. London Ser. A \textbf{392} (1984), no.~1802, 45--57.

\bibitem{bogo-shir2}
N.N. Bogoliubov and D.V. Shirkov, \emph{Introduction to the {T}heory of
  {Q}uantized {F}ields}, {R}ussian ed., John Wiley \& Sons, New
  York-Chichester-Brisbane, 1980, Translation edited by Seweryn Chomet, A
  Wiley-Interscience Publication.

\bibitem{dirac3}
P.A.M. Dirac, \emph{Discussion of the infinite distribution of electrons in the
  theory of the positron}, Proc. Camb. Philos. Soc. \textbf{30} (1934),
  150--163.

\bibitem{feynmanloop}
R.P. Feynman, \emph{Space-time approach to quantum electrodynamics}, Physical
  Rev. (2) \textbf{76} (1949), 769--789.

\bibitem{sea}
F.~Finster, \emph{Definition of the {D}irac sea in the presence of external
  fields}, arXiv:hep-th/9705006, Adv. Theor. Math. Phys. \textbf{2} (1998),
  no.~5, 963--985.

\bibitem{firstorder}
\bysame, \emph{Light-cone expansion of the {D}irac sea to first order in the
  external potential}, arXiv:hep-th/9707128, Michigan Math. J. \textbf{46}
  (1999), no.~2, 377--408.

\bibitem{light}
\bysame, \emph{Light-cone expansion of the {D}irac sea in the presence of
  chiral and scalar potentials}, arXiv:hep-th/9809019, J. Math. Phys.
  \textbf{41} (2000), no.~10, 6689--6746.

\bibitem{PFP}
\bysame, \emph{The {P}rinciple of the {F}ermionic {P}rojector}, hep-th/0001048,
  hep-th/0202059, hep-th/0210121, AMS/IP Studies in Advanced Mathematics,
  vol.~35, American Mathematical Society, Providence, RI, 2006.

\bibitem{sector}
\bysame, \emph{An action principle for an interacting fermion system and its
  analysis in the continuum limit}, arXiv:0908.1542 [math-ph] (2009).

\bibitem{srev}
\bysame, \emph{A formulation of quantum field theory realizing a sea of
  interacting {D}irac particles}, arXiv:0911.2102 [hep-th], Lett. Math. Phys.
  \textbf{97} (2011), no.~2, 165--183.

\bibitem{qft}
\bysame, \emph{Perturbative quantum field theory in the framework of the
  fermionic projector}, arXiv:1310.4121 [math-ph], J. Math. Phys. \textbf{55}
  (2014), no.~4, 042301.

\bibitem{grotz}
F.~Finster and A.~Grotz, \emph{The causal perturbation expansion revisited:
  Rescaling the interacting {D}irac sea}, arXiv:0901.0334 [math-ph], J. Math.
  Phys. \textbf{51} (2010), 072301.

\bibitem{rrev}
F.~Finster, A.~Grotz, and D.~Schiefeneder, \emph{Causal fermion systems: A
  quantum space-time emerging from an action principle}, arXiv:1102.2585
  [math-ph], Quantum {F}ield {T}heory and {G}ravity (F.~Finster, O.~M\"uller,
  M.~Nardmann, J.~Tolksdorf, and E.~Zeidler, eds.), Birkh\"auser Verlag, Basel,
  2012, pp.~157--182.

\bibitem{intro}
F.~Finster, J.~Kleiner, and J.-H. Treude, \emph{An introduction to the
  fermionic projector and causal fermion systems}, in preparation.

\bibitem{finite}
F.~Finster and M.~Reintjes, \emph{A non-perturbative construction of the
  fermionic projector on globally hyperbolic manifolds {I} -- {S}pace-times of
  finite lifetime}, arXiv:1301.5420 [math-ph] (2013).

\bibitem{infinite}
\bysame, \emph{A non-perturbative construction of the fermionic projector on
  globally hyperbolic manifolds {II} -- {S}pace-times of infinite lifetime},
  arXiv:1312.7209 [math-ph] (2013).

\bibitem{hadamard}
F.~Finster~et al, \emph{The fermionic projector in an external potential:
  Non-perturbative construction and the {H}adamard property}, in preparation.

\bibitem{heisenberg2}
W.~Heisenberg, \emph{Bemerkungen zur {D}iracschen {T}heorie des {P}ositrons},
  Z. Phys. \textbf{90} (1934), 209--231.

\bibitem{uehling}
E.A. Uehling, \emph{Polarization effects in the positron theory}, Phys. Rev.
  \textbf{48} (1935), 55--63.

\end{thebibliography}
\providecommand{\bysame}{\leavevmode\hbox to3em{\hrulefill}\thinspace}
\providecommand{\MR}{\relax\ifhmode\unskip\space\fi MR }
\providecommand{\MRhref}[2]{%
  \href{http://www.ams.org/mathscinet-getitem?mr=#1}{#2}
}
\providecommand{\href}[2]{#2}

\end{document}